\documentclass[12pt,letterpaper]{article}
\pdfoutput=1
\usepackage[pdftex]{graphicx,color}
\usepackage{hyperref}
\input{psfig}
\usepackage{amsmath,amssymb}
\usepackage[dvips,letterpaper,text={6.5in,9in}]{geometry}
\usepackage{fancyhdr}
\usepackage{verbatim}

\newcommand\ltap{\
  \raise.3ex\hbox{$<$\kern-.75em\lower1ex\hbox{$\sim$}}\ }
\newcommand\gtap{\
  \raise.3ex\hbox{$>$\kern-.75em\lower1ex\hbox{$\sim$}}\ }

\newcommand\simge{\mathrel{%
   \rlap{\raise 0.511ex \hbox{$>$}}{\lower 0.511ex \hbox{$\sim$}}}}
\newcommand\simle{\mathrel{
   \rlap{\raise 0.511ex \hbox{$<$}}{\lower 0.511ex \hbox{$\sim$}}}}

\newcommand{\slashchar}[1]%
        {\kern .25em\raise.18ex\hbox{$/$}\kern-.75em #1}
\def\lsim{\mathrel{\raise.3ex\hbox{$<$\kern-.75em\lower1ex\hbox{$\sim$}}}}
\def\gsim{\mathrel{\raise.3ex\hbox{$>$\kern-.75em\lower1ex\hbox{$\sim$}}}}
\newcommand{\bs}{\boldsymbol}

\newcommand\CL{{\cal L}}

\newcommand\CO{{\cal O}}

\newcommand\be{\begin{equation}}
\newcommand\ee{\end{equation}}
\newcommand\bea{\begin{eqnarray}}
\newcommand\eea{\end{eqnarray}}
\newcommand\ba{\begin{array}}
\newcommand\ea{\end{array}}
\newcommand\nn{\nonumber}

\newcommand{\thalf}{\textstyle{\frac{1}{2}}}
\newcommand{\tthird}{\textstyle{\frac{1}{3}}}
\newcommand{\tfourth}{\textstyle{\frac{1}{4}}}
\newcommand{\tfifth}{\textstyle{\frac{1}{5}}}

\newcommand\ra{\rightarrow}

\newcommand\gev{{\rm GeV}}
\newcommand\tev{{\rm TeV}}

\newcommand\pb{{\rm pb}}

\newcommand\fb{{\rm fb}}
\newcommand\ifb{{\rm fb}^{-1}}

\newcommand\etmiss{\slashchar{E}_T}

\newcommand\ellm{\ell^-}
\newcommand\ellpm{\ell^\pm}
\newcommand\ellp{\ell^+}

\newcommand\atc{\alpha_{TC}}

\newcommand\Ltc{\Lambda_{TC}}

\newcommand\grpp{g_{\rho_T\pi_T\pi_T}}

\newcommand\tom{\omega_{T}}
\newcommand\tro{\rho_{T}}
\newcommand\atro{\alpha_{\rho_T}}

\newcommand\ta{a_T}

\newcommand\tapm{a_T^\pm}

\newcommand\tropm{\rho_{T}^\pm}

\newcommand\tpi{\pi_T}
\newcommand\tpipm{\pi_T^\pm}

\newcommand\tpip{\pi_T^+}

\newcommand\tpiz{\pi_T^0}

\newcommand\jets{{\rm jets}}

\newcommand\Mjj{M_{jj}}

\newcommand\MWjj{M_{Wjj}}
\newcommand\MZjj{M_{Zjj}}
\newcommand\ptjj{p_{T}(jj)}

\newcommand\dRll{\Delta R_{\ell\ell}}
\newcommand\dXll{\Delta\chi_{\ell\ell}}

\newcommand\dphi{\Delta\phi}
\newcommand\deta{\Delta\eta}
\newcommand\dR{\Delta R}
\newcommand\dX{\Delta\chi}
\newcommand\dRm{(\Delta R)_{\rm min}}
\newcommand\dXm{(\Delta\chi)_{\rm min}}

\newcommand\cth{c_{\theta}}
\newcommand\sth{s_{\theta}}
\newcommand\cthst{c_{\theta^*}}
\newcommand\sthst{s_{\theta^*}}
\newcommand\cphst{c_{\phi^*}}
\newcommand\sphst{s_{\phi^*}}
\newcommand\bth{b_{\theta}}
\newcommand\bthst{b_{\theta^*}}
\newcommand\bphst{b_{\phi^*}}

\begin{document}

\title{
\vskip -15mm
\begin{flushright}
 \vskip -15mm
 {\small FERMILAB-Pub-12-253-T\\
   LAPTh-025/12\\
 }
 \vskip 5mm
 \end{flushright}
{\Large{\bf Testing the Technicolor Interpretation \\of the CDF Dijet
    Excess at the 8-TeV LHC}}\\
} \author{
  {\large Estia Eichten$^{1}$\thanks{eichten@fnal.gov} ,\,
  Kenneth Lane$^{2}$\thanks{lane@physics.bu.edu},\,
  Adam Martin$^{1}$\thanks{aomartin@fnal.gov}}\, and 
  Eric Pilon$^{3}$\thanks{pilon@lapp.in2p3.fr}\\
{\large {$^{1}$}Theoretical Physics Group, Fermi National Accelerator
  Laboratory}\\
{\large P.O. Box 500, Batavia, Illinois 60510}\\
{\large $^{2}$Department of Physics, Boston University}\\
{\large 590 Commonwealth Avenue, Boston, Massachusetts 02215}\\
{\large $^{3}$Laboratoire d'Annecy-le-Vieux de Physique Th\'eorique} \\
{\large UMR5108\,, Universit\'e de Savoie, CNRS} \\
{\large B.P.~110, F-74941, Annecy-le-Vieux Cedex, France}\\
}
\maketitle

\begin{abstract}
  
  Under the assumption that the dijet excess seen by the CDF Collaboration
  near $150\,\gev$ in $Wjj$ production is due to the lightest technipion of
  the low-scale technicolor process $\tro \ra W\tpi$, we study its
  observability in LHC detectors for $\sqrt{s} = 8\,\tev$ and $\int\CL dt =
  20\,\ifb$. We describe interesting new kinematic tests that can provide
  independent confirmation of this LSTC hypothesis. We show that cuts similar
  to those employed by CDF, and recently by ATLAS, cannot confirm the dijet
  signal. We propose cuts tailored to the LSTC hypothesis and its backgrounds
  at the LHC that may reveal $\tro \ra \ell\nu jj$. Observation of the
  isospin-related channel $\tropm \ra Z\tpipm \ra \ellp\ellm jj$ and of
  $\tropm \ra WZ$ in the $\ellp\ellm\ellpm\nu_\ell$ and $\ellp\ellm jj$ modes
  will be important confirmations of the LSTC interpretation of the CDF
  signal. The $Z\tpi$ channel is experimentally cleaner than $W\tpi$ and its
  rate is known from $W\tpi$ by phase space. It can be discovered or excluded
  with the collider data expected by the end of 2012. The $WZ \ra 3\ell\nu$
  channel is cleanest of all and its rate is determined from $W\tpi$ and the
  LSTC parameter $\sin\chi$. This channel and $WZ \to \ellp\ellm jj$ are
  discussed as a function of $\sin\chi$.

\end{abstract}


\newpage

\section*{1. Introduction}

The CDF Collaboration has reported evidence for a resonance near $150\,\gev$
in the dijet-mass spectrum, $\Mjj$, of $Wjj$ production. This was based on an
integrated luminosity of $4.3\,\ifb$~\cite{Aaltonen:2011mk} and updated with
a total data sample of $7.3\,\ifb$~\cite{CDFnew}. In Ref.~\cite{CDFnew}, the
resonant dijet excess has a significance of $4.1\,\sigma$.  The D\O\
Collaboration, on the other hand, published a search for this resonance based
on $4.3\,\ifb$ that found no significant excess. Based on a $W+$Higgs boson
production model, D\O\ determined a cross section for a potential signal of
$0.82^{+ 0.83}_{- 0.82}\,\pb$ and a 95\% confidence level upper limit of
$1.9\,\pb$~\cite{Abazov:2011af}. Analyzing its data with the same production
model, CDF reported a signal rate of $3.0\pm 0.7\,\pb$ and a discrepancy
between the two experiments of $2.5\,\sigma$~\cite{AnnoviLP11}. This
discrepancy remains. The purpose of this paper is to help guide the LHC
experiments in searches to test for the CDF dijet excess in the $Wjj$ and two
closely related channels. We do this in the context low-scale technicolor
(LSTC), interpreting CDF's dijet excess as the lightest technipion
$\pi_T^{\pm,0}$ of this scenario, produced in association with $W^\pm$ in the
decay $\rho_T^{\pm,0}, a_T^{\pm,0} \ra W\tpi$ and decaying to a pair of quark
jets~\cite{Eichten:2011sh}. The related channels supporting this
interpretation are $\tropm,\tapm \ra Z\tpipm$ and $W^\pm Z$.\footnote{LHC
  studies of the $Wjj$ and $WZ$ channels carried out so far are discussed in
  Secs.~3 and ~5, respectively.} They require {\em no} additional LSTC model
assumptions beyond those made in Ref.~\cite{Eichten:2011sh} to determine LHC
production rates.  We assume $\sqrt{s} = 8\,\tev$ and consider $\int\CL dt =
20\,\ifb$, the amount of data expected to be in hand by the end of
2012.\footnote{Preliminary versions of this paper were circulated in
  Ref.~\cite{Eichten:2011xd} and Ref.~\cite{Eichten:2012br} assuming
  $\sqrt{s} = 7\,\tev$ and $\int\CL dt = 1$--$20\,\ifb$. The simulations in
  the current paper may be applied to different luminosities by scaling the
  event rates. We have not included the nontrivial effects of pileup at the
  higher luminosities of 8-TeV running. They also make difficult a detailed
  comparison of our results with the earlier 7-TeV ones. Our signal cross
  sections are uniformly 20\% greater at $8\,\tev$ than at $7\,\tev$, but the
  increases in various physics backgrounds are not so simply summarized.}

Low-scale technicolor (LSTC) is a phenomenology based on walking
technicolor~\cite{Holdom:1981rm, Appelquist:1986an,Yamawaki:1986zg,
  Akiba:1986rr}. The gauge coupling $\atc$ must run very slowly for 100s of
TeV above the TC scale, $\Ltc \sim$ several $100\,\gev$, so that extended
technicolor (ETC) can generate sizable quark and lepton masses while
suppressing flavor-changing neutral current
interactions~\cite{Eichten:1979ah}. This may be achieved, e.g., with
technifermions belonging to higher-dimensional representations of the TC
gauge group. Then, the constraints of Ref.~\cite{Eichten:1979ah} on the
number of ETC-fermion representations imply that there will be technifermions
in the fundamental TC representation as well. They are expected to condense
at an appreciably lower energy scale than those belonging to the
higher-dimensional representations and, thus, their technipions' decay
constant $F_1^2 \ll F_\pi^2 = (246\,\gev)^2$~\cite{Lane:1989ej}. Spin-one
bound states of these technifermions will have an orthoquarkonium-like
spectrum with masses well below a TeV --- greater than the previous Tevatron
limit $M_{\tro} \simge 250\,\gev$~\cite{Abazov:2006iq, Aaltonen:2009jb} and
probably less than 600--700~GeV, a scale at which we believe the notion of
``low-scale'' TC ceases to make sense. The most accessible states are the
lightest technivectors, $V_T = \tro(I^G J^{PC} = 1^+1^{--})$,
$\tom(0^-1^{--})$ and $\ta(1^-1^{++})$. Through their mixing with the
electroweak bosons, they are readily produced as $s$-channel resonances via
the Drell-Yan process in colliders. Spin-zero technipions $\tpi(1^-0^{-+})$
are accessed in $V_T$ decays. A central assumption of LSTC is that these
lightest technihadrons may be treated in isolation, without significant
mixing or other interference from higher-mass states.  Also, we expect that
(1) the lightest technifermions are $SU(3)$-color singlets, (2) isospin
violation is small for $V_T$ and $\tpi$, (3) $M_{\tom} \cong M_{\tro}$, and
(4) $M_{\ta}$ is not far above $M_{\tro}$.  This last assumption is made to
keep the low-scale TC contribution to the $S$-parameter small. An extensive
discussion of LSTC, including these points and precision electroweak
constraints, is given in Ref.~\cite{Lane:2009ct}.

Walking technicolor has another important consequence: it enhances $M_{\tpi}$
relative to $M_{\tro}$ so that the all-$\tpi$ decay channels of the $V_T$ are
likely to be closed~\cite{Lane:1989ej}. Principal $V_T$-decay modes are
$W\tpi$, $Z\tpi$, $\gamma \tpi$, a pair of EW bosons (which can include one
photon), and fermion-antifermion pairs~\cite{Lane:2002sm,
  Eichten:2007sx,Lane:2009ct}. If allowed by isospin, parity and angular
momentum, $V_T$ decays to one or more weak bosons involve
longitudinally-polarized $W_L/Z_L$, the technipions absorbed via the Higgs
mechanism. The rates for these nominally strong decays are suppressed by
powers of $\sin^2\chi = (F_1/F_\pi)^2 \ll 1$. This important LSTC parameter
is a mixing factor that measures the amount that the lowest-scale technipion
is the mass eigenstate $\tpi$ ($\cos\chi$) and the amount that it is
$W_L/Z_L$ ($\sin\chi$). Thus, each replacement of a mass-eigenstate $\tpi$ by
$W_L/Z_L$ in a $V_T$ decay amplitude costs a factor of $\tan\chi$. Decays to
transversely-polarized $\gamma,W_\perp,Z_\perp$ are suppressed by $g,g'$.
Thus, the $V_T$ are {\em very} narrow, $\Gamma(\tro) \simle 1\,\gev$ and
$\Gamma(\tom,\ta) \simle 0.1\,\gev$ for the masses considered here. These
decays have striking signatures, visible above backgrounds within a limited
mass range at the Tevatron and probably up to 600--700~GeV at the
LHC~\cite{Brooijmans:2008se, Brooijmans:2010tn}.

In Ref.~\cite{Eichten:2011sh} we proposed that CDF's dijet excess is due to
resonant production of $W\tpi$ with $M_{\tpi} = 160\,\gev$. We took $M_{\tro}
= 290\,\gev$ and $M_{\ta} = 1.1 M_{\tro} = 320\,\gev$.\footnote{The Pythia
  default decays for technipions are based on the assumption that they are
  Higgs-like, i.e., involve couplings proportional to fermion mass. They are
  thus dominated by $\tpip \to c \bar b$, $u \bar b$ and $\tpiz \to b \bar
  b$. These modes involve energy loss to neutrinos that we have not included
  in reconstructing dijet masses. Therefore, the choice $M_{\tpi} =
  160\,\gev$ reconstructs close to $150\,\gev$. {\em If} technipions decay
  mainly to light quarks and leptons, a plausible possibility for the
  lightest $\tpi$, then we would expect all our input technihadron masses to
  decrease by 10--$15\,\gev$.} Then, about 75\% of the $W\tpi$ rate at the
Tevatron is due to $\tro \ra W\tpi$ and, of this, most of the $W$'s are
longitudinally polarized.\footnote{About 70\% of the $W\tpi$ rate at the LHC
  is due to the $\tro$.} The remainder is dominated by $\ta$ production. Its
decay, and a small fraction of the $\tro$'s, involve $W_\perp$ production,
which is generated by dimension-five operators~\cite{Lane:2009ct}. These
operators are suppressed by mass parameters $M_{V,A}$ that we take equal to
$M_{\tro}$. The other LSTC parameters relevant to $W\tpi$ production are
$\grpp$ and $\sin\chi$. The $\tro \to \tpi\tpi$ coupling $\grpp$ is the same
for all $\tro$ decays considered here and it is naively scaled from QCD; its
{\sc Pythia} default value is $\atro = \grpp^2/4\pi = 2.16(3/N_{TC})$ with
$N_{TC} = 4$. We use $\sin\chi = 1/3$. Using the LSTC model implemented in
{\sc Pythia}~\cite{Lane:2002sm,Eichten:2007sx, Sjostrand:2006za}, we found
$\sigma(\bar pp \ra \tro \ra W\tpi \ra Wjj) = 2.2\,\pb$ ($480\,\fb$ for $W
\ra e\nu,\,\mu\nu$).\footnote{This includes $B(\tpi \ra \bar q q) \simeq
  90\%$ in the default {\sc Pythia} $\tpi$-decay table.} Adopting CDF's cuts,
we closely matched its $\Mjj$ distribution for signal and
background. Motivated by the peculiar kinematics of $\tro$ production at the
Tevatron and $\tro \ra W\tpi$ decay, we also suggested cuts intended to
enhance the $\tpi$ signal's significance and to make $\tro \ra Wjj$
visible. Several distributions of data in the excess region $115\,\gev < \Mjj
< 175\,\gev$ published by CDF~\cite{CDFnew} --- notably $\MWjj$, $\ptjj$,
$\dphi$ and $\dR = \sqrt{(\deta)^2 + (\dphi)^2}$ --- fit the expectations of
the LSTC model very well. The background-subtracted $\dR$ distribution, in
particular, has a behavior which, we believe, furnishes strong support for
our dijet production mechanism.

The purpose of this paper is to propose and study ways to test for the CDF
signal at the LHC. In Sec.~2 we review the kinematics of $\tro, \ta \ra
W\tpi$ and $Z\tpi$ in LSTC. We also present an interesting new result: the
nonanalytic behavior of $d\sigma/d(\dR$) and $d\sigma/d(\dX)$ at their
thresholds, $\dRm$ and $\dXm$. Here $\dX$ is the opening angle between the
$\tpi$ decay jets in the $\tro$ rest frame. For massless jets, a good
approximation, we find that $\dRm = \dXm = 2\cos^{-1}(v)$, where $v =
p_{\tpi}/E_{\tpi}$ is the $\tpi$ velocity in the $\tro$ rest frame. This
result, peculiar to production models such as LSTC in which a narrow
resonance decays to another narrow resonance plus a $W$ or $Z$, provides
measures of $v$ independent of $p/E$ and, hence, valuable corroboration of
this type of production. In Sec.~3 we consider the $\tro,\ta \ra W\tpi$
process. Its LHC cross section at $8\,\tev$ is $9.5\,\pb$ but, for CDF cuts,
its backgrounds have increased by about a factor of ten over those at the
Tevatron. This makes testing for the dijet excess in this channel very
challenging. We suggest cuts which enhance signal-to-background $(S/B)$ but
which will still require a very good understanding of the backgrounds in
$Wjj$ production. Recent studies of $Wjj$ production by ATLAS and CMS are
discussed there. In Sec.~4 we study $\tropm,\tapm \ra Z\tpipm$, whose cross
section is $2.8\,\pb$ at $8\,\tev$ ($190\,\fb$ after $Z \ra e^+e^-,\,
\mu^+\mu^-$). This is the isospin partner of $\tropm,\tapm \ra W\tpiz$, so
its cross section is rather confidently known. The $\ellp\ellm jj$ channel is
free of QCD multijet and $\bar t t$ backgrounds and missing energy
uncertainty. Reconstructing the $Zjj$ invariant mass and other signal
distributions, particularly in $\dR$ and $\dX$, will benefit from this. {\em
  Because of these features, we believe that the $Z\tpi \ra Zjj$ mode will be
  the surest test of CDF's dijet signal at the LHC.} In Sec.~5, we study
$\tropm,\tapm \ra WZ$. The cross section for this mode is proportional to
$\tan^2\chi$ times the $\tropm,\tapm \ra W^\pm\tpiz$ and $Z\tpipm$ rates, but
enhanced by its greater phase space. We predict $\sigma(\tropm,\tapm \ra WZ)
= 1.8\,\ (1.1)\,\pb$ for $\sin\chi = 1/3\, (1/4)$. In the all-leptons
$3\ell\nu$ mode with $e$'s and $\mu$'s, the rate is only $26\, (15)\,\fb$,
but jet-related uncertainties are absent except insofar as they effect
$\etmiss$ resolution. A new study by CMS of this channel is discussed
there. The $WZ \to \ellp\ellm jj$ mode is also an interesting target of
opportunity so long as $\sin\chi \simge 1/4$. The $\dR$ and $\dX$
distributions for $Z \to jj$ again provide support for our narrow
LSTC-resonance production model. In short, one or both of the $Z\tpi$ and
$WZ$ modes should be dispositive of the LSTC interpretation of the CDF dijet
excess with the $\sim 20\,\ifb$ expected by the end of 2012. We present in an
appendix the details of calculations in Sec.~2 regarding the nonanalytic
threshold behavior of the $\dX$ and $\dR$
distributions. 

While the simulations of the CDF signal in this paper are made in the context
of low-scale technicolor, their qualitative features apply to any model in
which that signal is due to $\bar q q$ production of a narrow resonance
decaying to a $W$ plus another narrow resonance. Several papers have appeared
proposing such an $s$-channel mechanism~\cite{Kilic:2011sr,Cao:2011yt,
  Chen:2011wp, Fan:2011vw, Ghosh:2011np,Gunion:2011bx}. With similar
resonance masses to our LSTC proposal, these models will have kinematic
distributions like those we describe in Sec.~2. However, not all these models
will have the $Zjj$ and $WZ$ signals of LSTC. There are also a large number
of papers proposing that the CDF signal is due to production of a new
particle (e.g., a leptophobic $Z'$) that is not resonantly
produced~\cite{Buckley:2011vc, Hewett:2011fk, Harnik:2011mv, Dobrescu:2011fk,
  Nelson:2011us, Yu:2011cw, Cheung:2011zt}. These ``$t$-channel'' models will
not pass our kinematic tests.

\section*{2. LSTC Kinematics and Threshold Nonanalyticity}

The kinematics of $\tro \ra W\tpi$ at the Tevatron and LHC are a consequence
of the basic LSTC feature that walking TC enhancements of $M_{\tpi}$ strongly
suggest $M_{\tro} < 2M_{\tpi}$ and, indeed, that the phase space for $\tro
\ra W\tpi$ is quite limited~\cite{Lane:1989ej, Eichten:1997yq}. At the
Tevatron, a $290\,\gev$ $\tro$ is produced almost at rest, with almost no
$p_T$ and very little boost along the beam direction. At the LHC, $p_T(\tro)
\simle 25\,\gev$ and $\eta(\tro) \simle 2.0$. Furthermore, the $\tpi$ is
emitted very slowly in the $\tro$ rest frame --- $v \simeq 0.4$ for our
assumed masses --- so that its decay jets are roughly back-to-back in the lab
frame. Thus, $p_T(\tpi) \simle 80\,\gev$ and the $z$-boost invariant
quantities $\dphi$ and $\dR = \sqrt{(\deta)^2 + (\dphi)^2}$ are peaked at
large values less than $\pi$.

These features of LSTC are supported by CDF's $7.3\,\ifb$
data~\cite{CDFnew}. Figures~\ref{fig:CDFMWjj}--\ref{fig:CDFdRjj} show
distributions before and after background subtraction taken from the $115 <
\Mjj < 175\,\gev$ region containing the dijet excess. The subtracted-data
$\MWjj$ signal has a narrow resonant shape quite near
$290\,\gev$. Unfortunately, the background peaks not far below that mass so
that one may be concerned that the subtracted data's peak is due to
underestimating the background. Also, as we expect, the subtracted $p_T(jj)$
data falls off sharply above $75\,\gev$ and the subtracted $\dphi$ data is
strongly peaked at large values. Again, one may worry that these are
artifacts of the peak of the $\MWjj$ background and the position of the
$\Mjj$ excess.

\begin{figure}[!t]
 \begin{center}
\includegraphics[width=3.15in, height=3.15in]{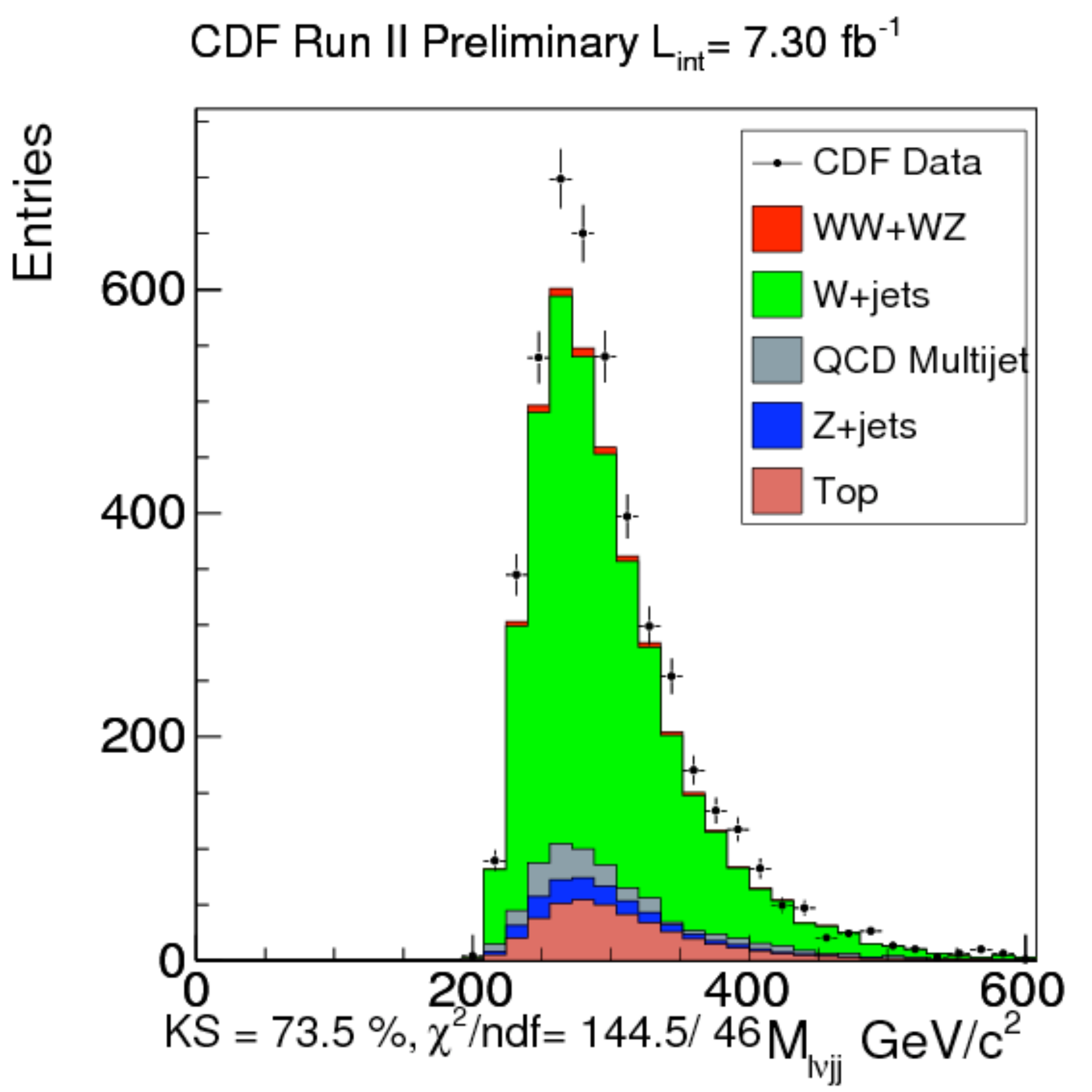}
\includegraphics[width=3.15in, height=3.15in]{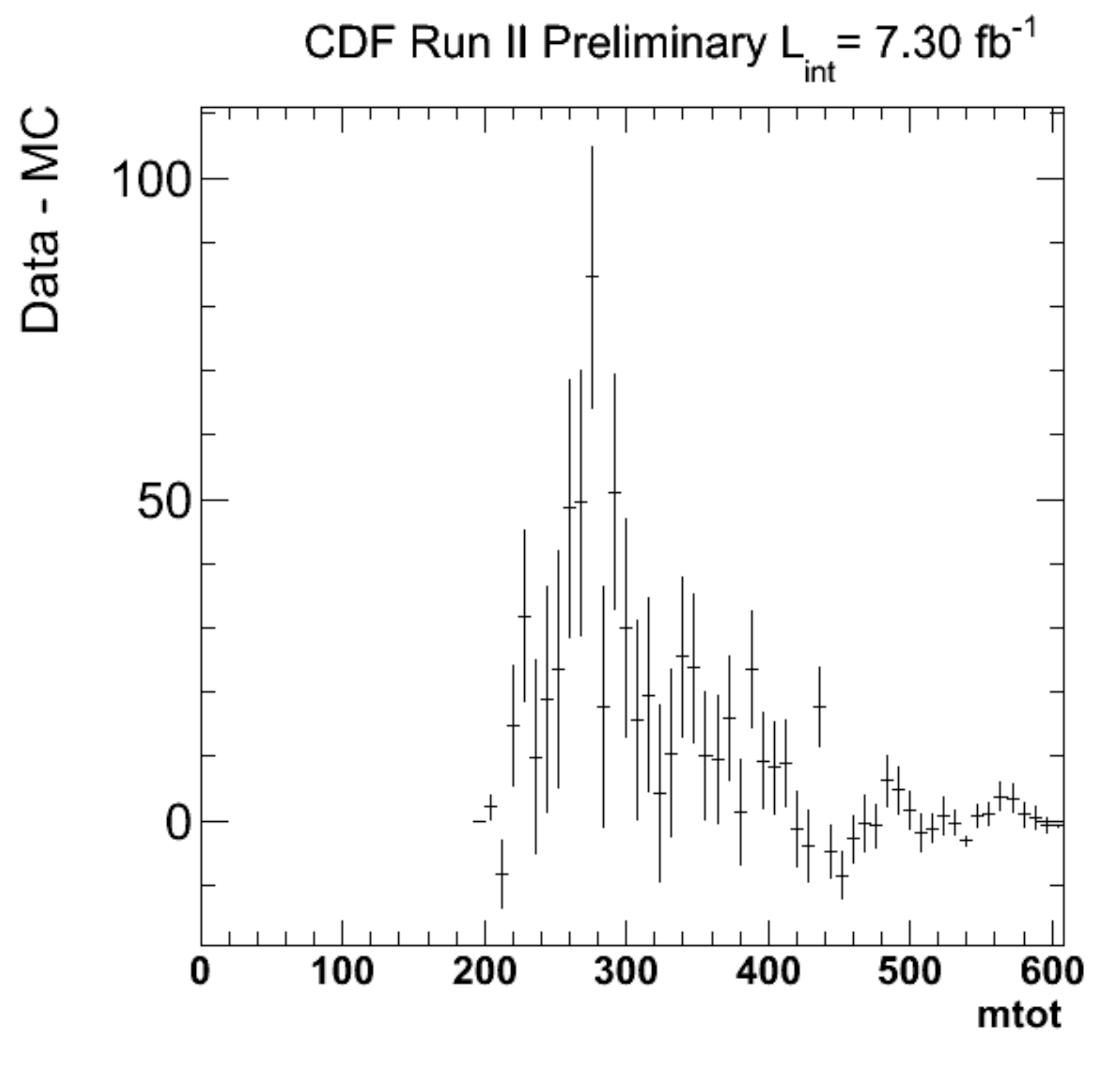}
\caption{CDF $\MWjj$ distributions for $\int \CL dt =
  7.3\,\ifb$ from the dijet signal region $115 < \Mjj <
  175\,\gev$~\cite{CDFnew}. Left: Expected backgrounds and data; right:
  background subtracted data.}
  \label{fig:CDFMWjj}
 \end{center}
 \end{figure}
\begin{figure}[!ht]
 \begin{center}
\includegraphics[width=3.15in, height=3.15in]{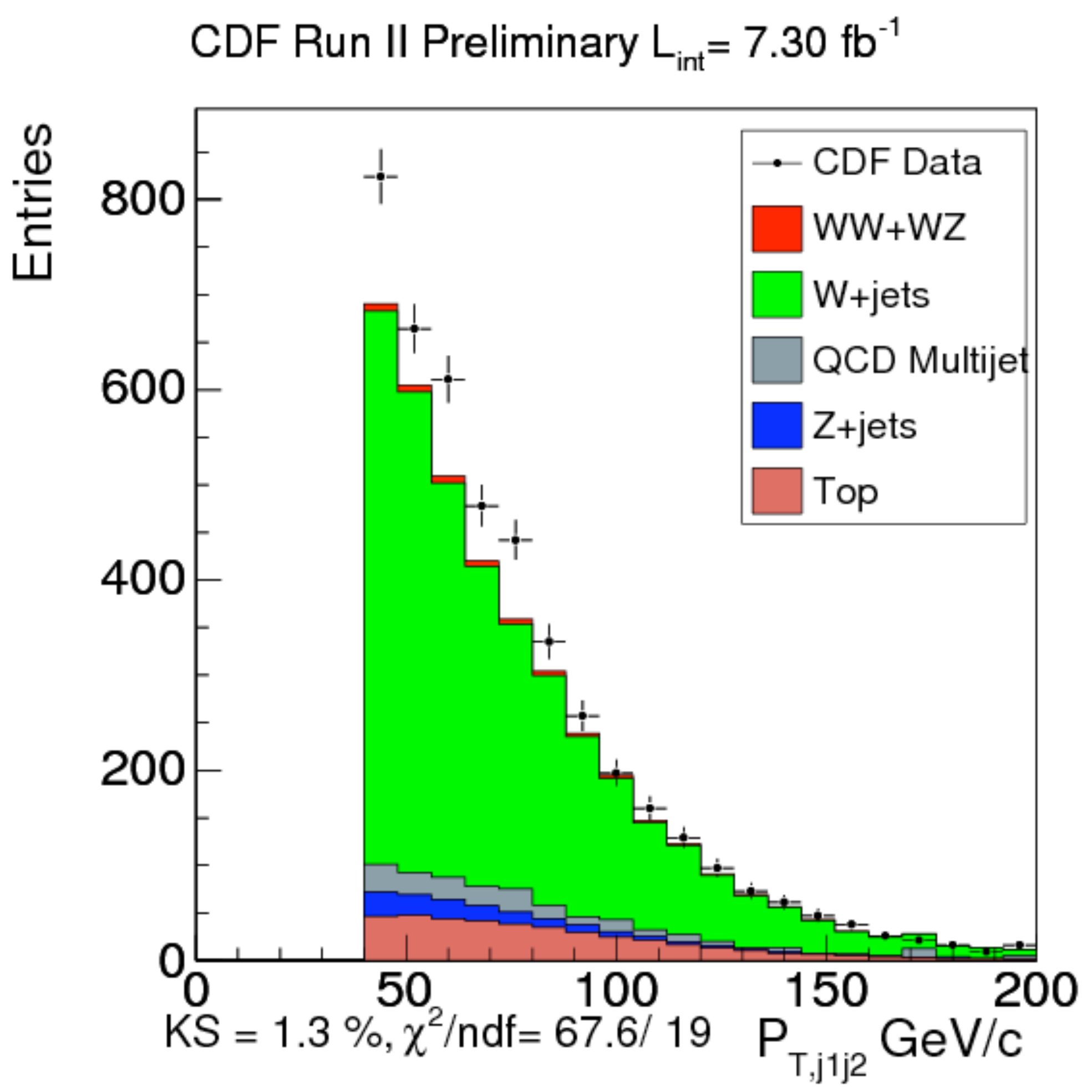}
\includegraphics[width=3.15in, height=3.15in]{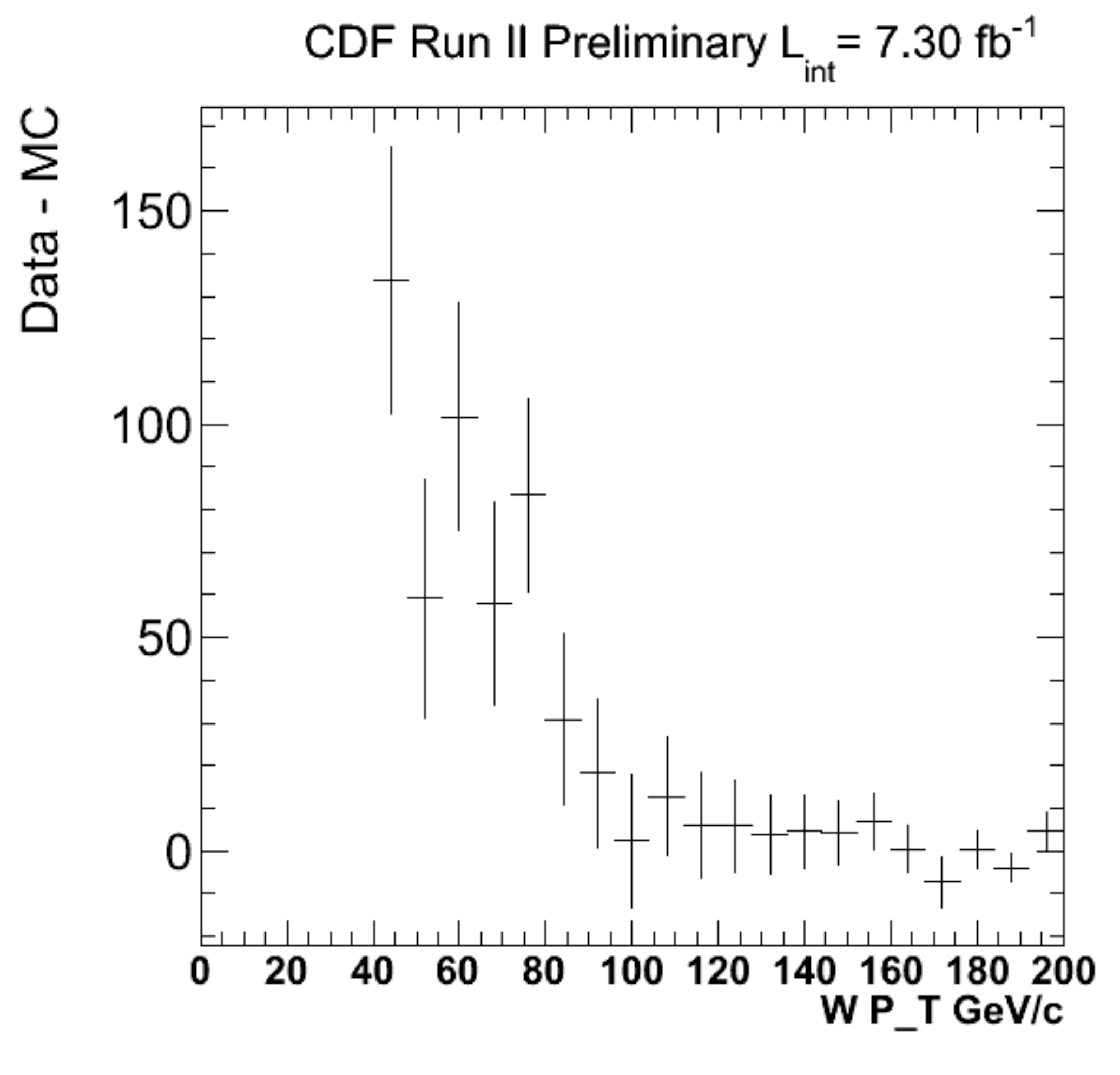}
\caption{CDF $p_T(jj)$ distributions for $\int \CL dt =
  7.3\,\ifb$ from the dijet signal region $115 < \Mjj <
  175\,\gev$~\cite{CDFnew}. Left: Expected backgrounds and data; right:
  background subtracted data.}
  \label{fig:CDFpTjj}
 \end{center}
 \end{figure}

 The background-subtracted $\dR$ distribution, however, is very interesting.
 It is practically zero for $\dR < 2.25$, then rises sharply to a broad
 maximum before falling to zero again at $\dR \simeq 3.5$. This behavior, and
 a somewhat similar one we predict for $\dX$ are the main subject of this
 section. We will show that the threshold form of the $\dR$ and $\dX$
 distributions provide direct measures of the velocity of the dijet system in
 the subprocess center-of-mass frame that are independent of measuring $p/E$
 and, thus, are independent checks on the two-resonance topology of the
 dijet's production mechanism.\footnote{Note that $\dR$ and $\dX$ are largely
   unaffected by lost neutrinos if semileptonic $b$-decays are an important
   component of $\tpi$ decays. Also, $\dX$ is defined in the $\tro$ rest
   frame, while $\dR$ is defined in the lab frame.  If one wishes to remove
   the effect of $p_T(\tro)$ on $\dR$, it should be defined in the $\tro$
   frame.}  One might think that the corresponding $\dRll$ and $\dXll$
 distributions from $Z \to \ellp\ellm$ would be similarly
 valuable. Unfortunately, because the dileptons come from real $Z$'s and our
 cuts make the background $Z$'s like the signal ones, $\dRll$ and $\dXll$ are
 indistinguishable from their backgrounds.

\begin{figure}[!t]
 \begin{center}
\includegraphics[width=3.15in, height=3.15in]{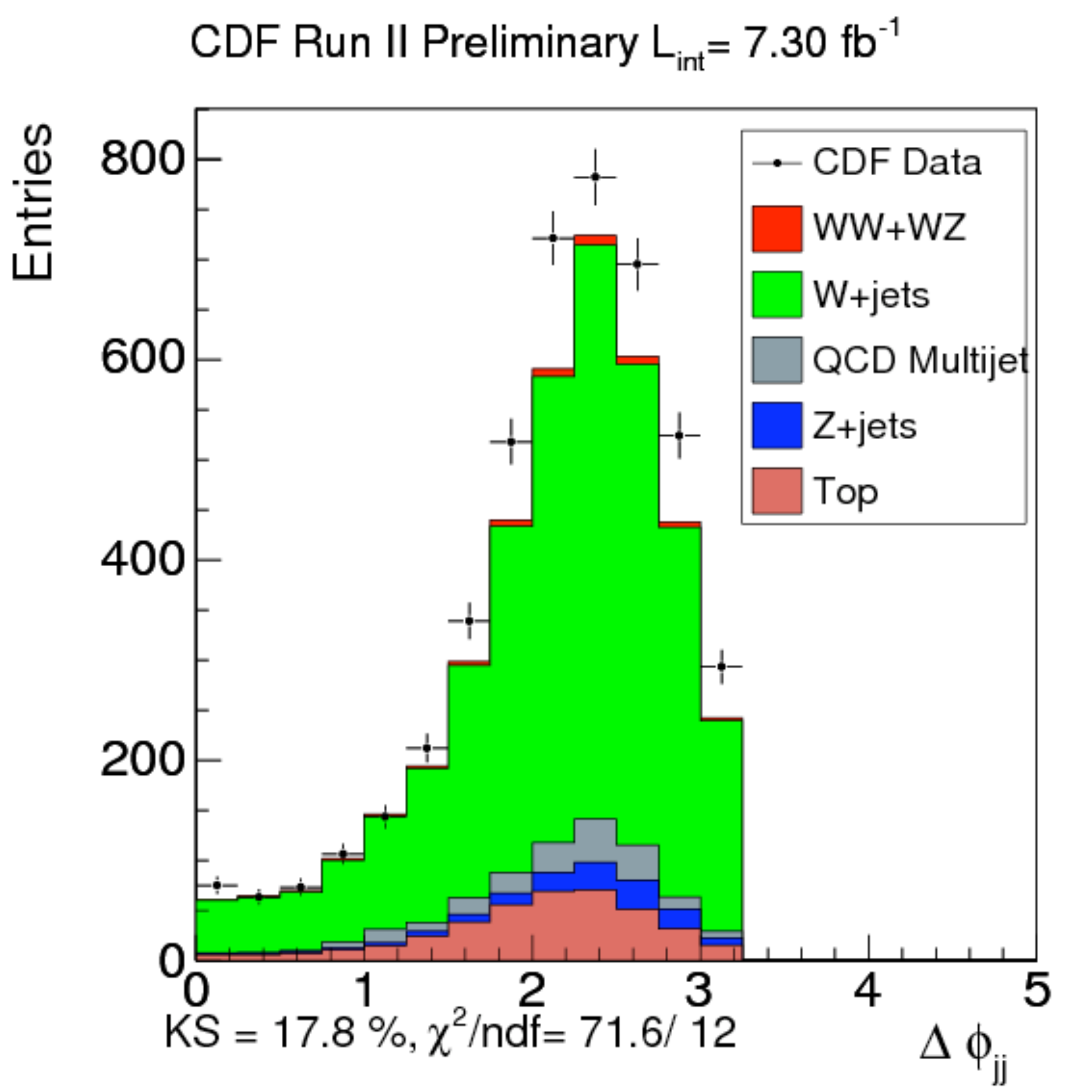}
\includegraphics[width=3.15in, height=3.15in]{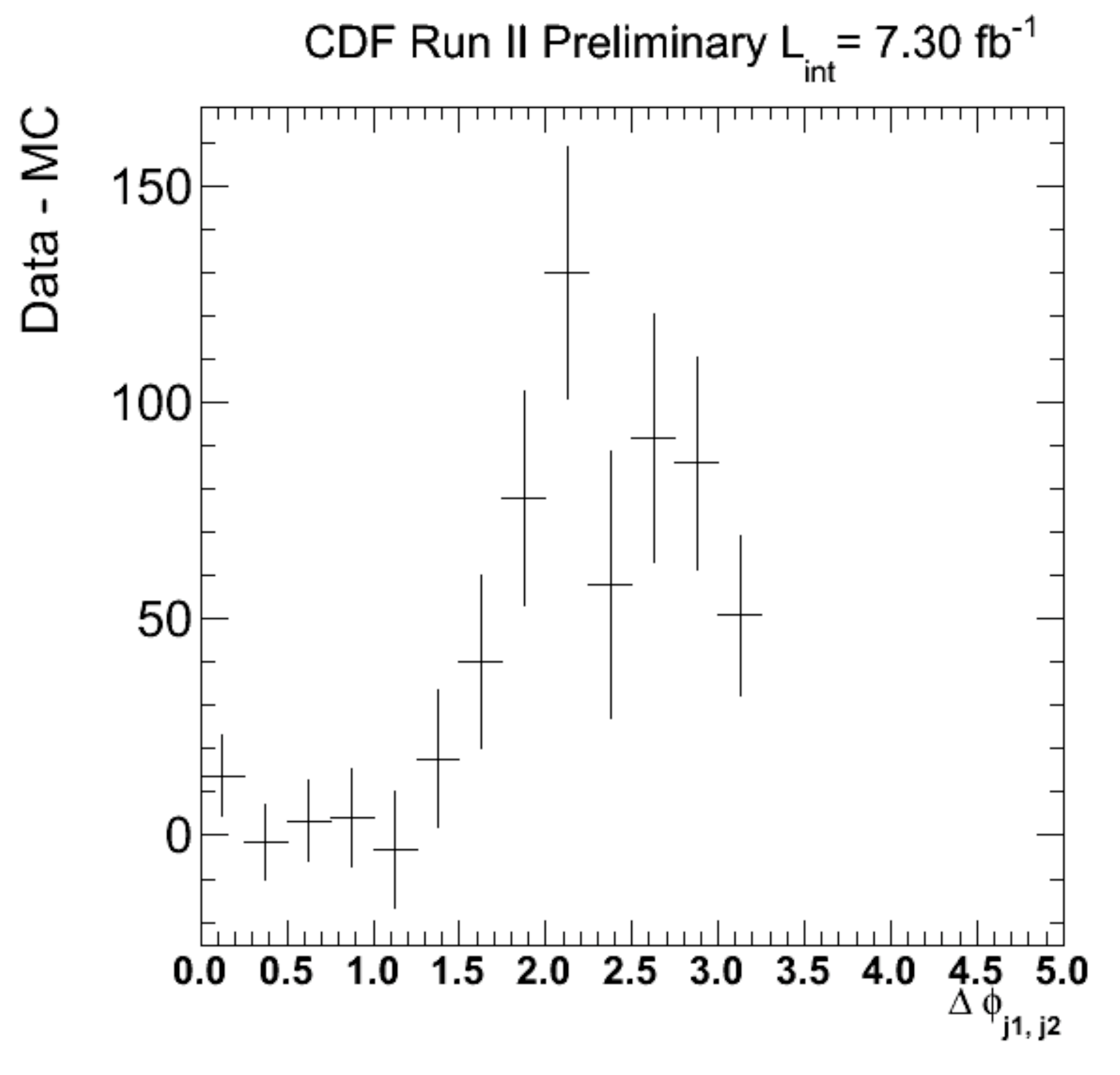}
\caption{CDF $\dphi$ distributions for $\int \CL dt =
  7.3\,\ifb$ from the dijet signal region $115 < \Mjj <
  175\,\gev$~\cite{CDFnew}. Left: Expected backgrounds and data; right:
  background subtracted data.}
  \label{fig:CDFdphijj}
 \end{center}
 \end{figure}
\begin{figure}[!ht]
 \begin{center}
\includegraphics[width=3.15in, height=3.15in]{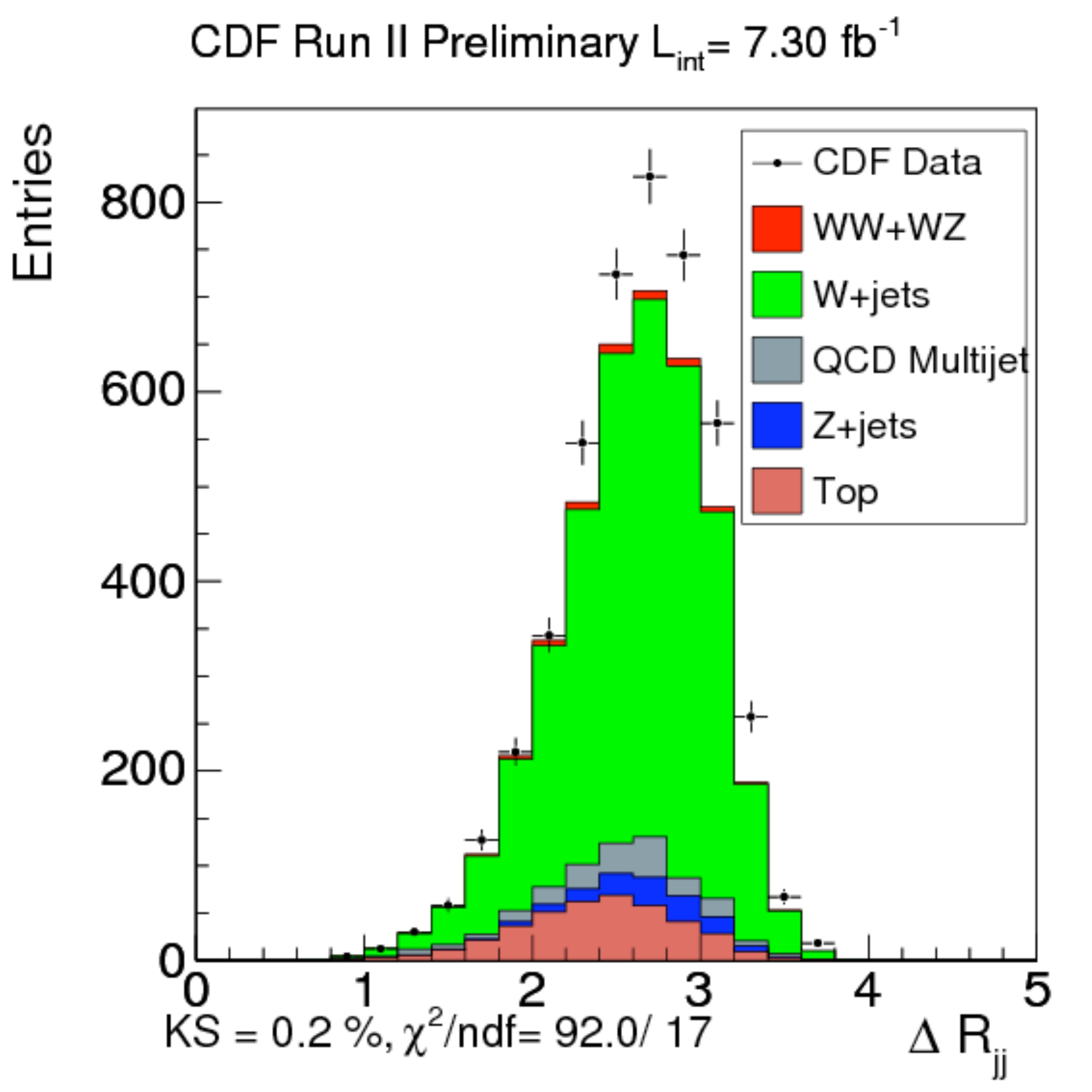}
\includegraphics[width=3.15in, height=3.15in]{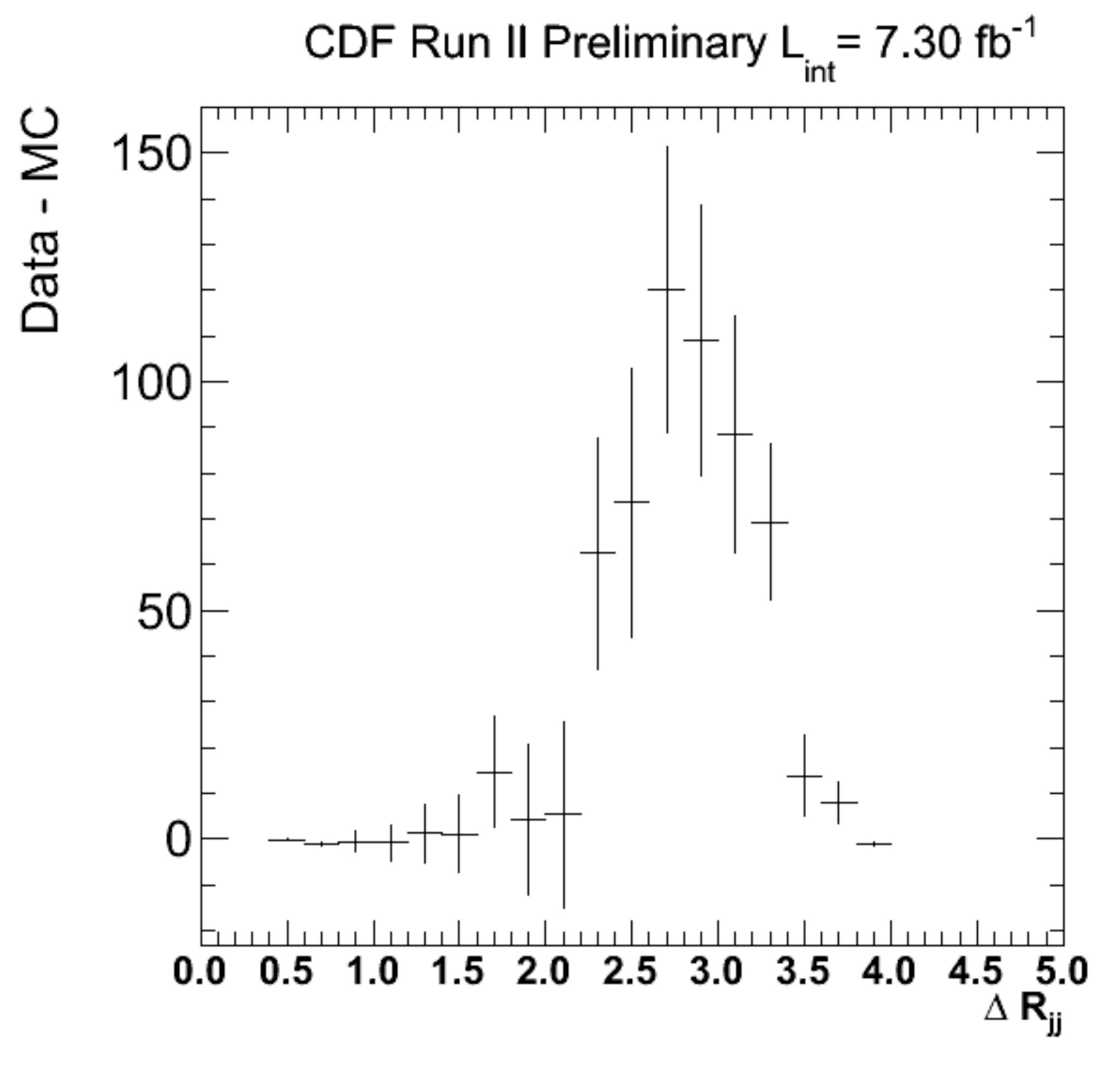}
\caption{CDF $\dR$ distributions for $\int \CL dt =
  7.3\,\ifb$ from the dijet signal region $115 < \Mjj <
  175\,\gev$~\cite{CDFnew}. Left: Expected backgrounds and data; right:
  background subtracted data.}
  \label{fig:CDFdRjj}
 \end{center}
 \end{figure}

 For our analysis, we assume the jets from $\tpi$ decay are massless. We have
 examined the effect of including jet masses and found them to be
 unimportant. We will remark briefly on this at the end of this section. We
 first consider the dominant $\tro$ contribution to $W/Z\tpi$ production,
 commenting on the $\ta$ contribution also at the end.
 
 Define the angles $\theta$, $\theta^*$ and $\phi^*$ as follows: Choose the
 $z$-axis as the direction of the event's boost; this is usually the
 direction of the incoming quark in the subprocess c.m.~frame. In the $\tro$
 rest frame, $\theta$ is the polar angle of the $\tpi$ velocity ${\bs v}$,
 the angle it makes with the $z$-axis.  Define the $xz$-plane as the one
 containing the unit vectors $\hat{\bs z}$ and $\hat{\bs v}$, so that
 $\hat{\bs v} = \hat{\bs x}\sin\theta + \hat{\bs z}\cos\theta$, and $\hat{\bs
   y} = \hat{\bs z} \times \hat{\bs x}$. Define a starred coordinate system
 {\em in the $\tpi$ rest frame} by making a rotation by angle~$\theta$ about
 the $y$-axis of the $\tro$ frame. This rotation takes $\hat{\bs z}$ into
 $\hat{\bs z}^* = \hat{\bs v}$ and $\hat{\bs x}$ into $\hat{\bs x}^* =
 \hat{\bs x}\cos\theta - \hat{\bs z}\sin\theta$. In this frame, let $\hat
 {\bs p}_1^*$ be the unit vector in the direction one of the jets (partons).
 The angle between $\hat{\bs v}$ and $\hat {\bs p}_1^*$ is $\theta^*$; the
 azimuthal angle of ${\bs p}_1^* = -{\bs p}_2^*$ is $\phi^*$:
\be\label{eq:angles}
\cos\theta = \hat{\bs z}\cdot \hat{\bs v}, \quad
\cos\theta^* = \hat{\bs p}_1^*\cdot \hat{\bs v}, \quad
\tan\phi^* = p_{1y^*}^*/p_{1x^*}^*.
\ee
Note that, since $\tpi \ra \bar q q$ is isotropic in its rest frame,
$d\sigma(\bar q q \ra \tro \ra Wjj)/d(\cos\theta^*) = \sigma/2$, where
$\sigma$ is the total subprocess cross section.

It is easier to consider the $d\sigma/d(\dX)$ distribution first. For
massless jets,
\be\label{eq:cosdchi}
1-\cos(\dX) = \frac{2(1-v^2)}{1-v^2\cos^2\theta^*}\,.
\ee
The minimum value of $\dX$ occurs when $\theta^* = \pi/2$ (i.e., ${\bs v}
\perp {\bs p}_1^*$), and so
\be\label{eq:delchimin}
\pi \ge \dX \ge \dXm = 2\cos^{-1}(v)\,.
\ee
From Eq.~(\ref{eq:cosdchi}), it is easy to see that
\be
\frac{d\sigma}{d(\dX)} = \frac{(1-v^2)\,\sigma}{4v\sin^2(\dX/2)
  \sqrt{\cos^2(\dXm/2) - \cos^2((\dX)/2)}}\,.
\ee
The $\dX$ distribution has an inverse-square-root singularity at $\dX = \dXm
= 2\cos^{-1}(v) = 2.23$ for our input masses, and falls sharply above
there. This is illustrated in Fig.~\ref{fig:dchidR} where we plot this
distribution for the primary partons and for the reconstructed jets. The
low-side tail for the jets is an artifact of their reconstruction.

\begin{figure}[!t]
 \begin{center}
\includegraphics[width=6.50in, height=3.15in]{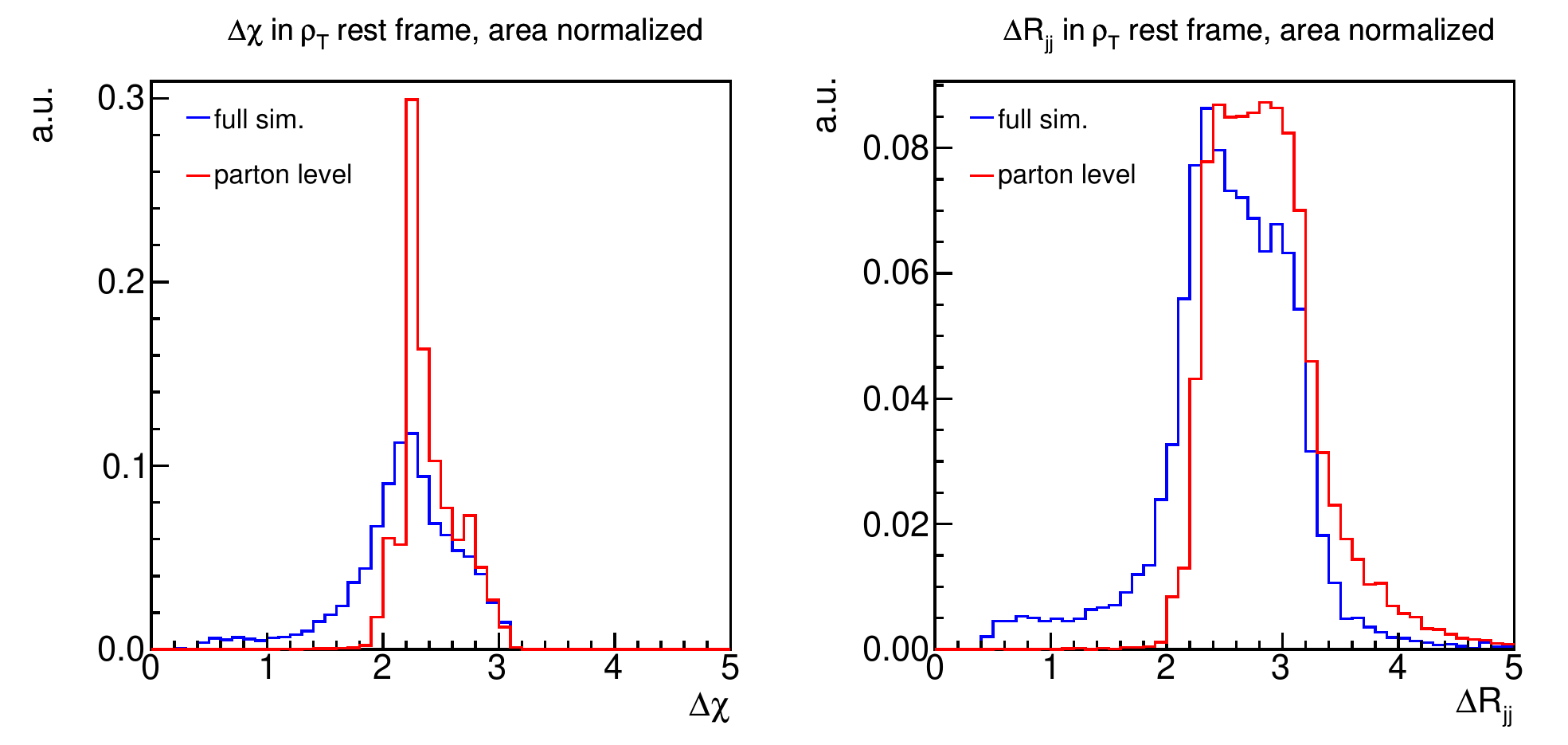}
\caption{The area-normalized $\dX$ and $\dR$ distributions for the primary
  parton/jet in $\tro,\ta \ra W\tpi$ production followed by $\tpi \ra \bar q
  q$ decay, constructed as described in the text. Red: pure distribution of
  primary parton before any radiation; blue: the distribution for the jets
  reconstructed as described in Sec.~3.}
  \label{fig:dchidR}
 \end{center}
 \end{figure}

To understand this singularity better, it follows from
Eq.~(\ref{eq:cosdchi}) that $\dX$ may be expanded about $\cos\theta^* = 0$
as
\be\label{eq:dXexpand}
\dX = \dXm + \frac{a}{2}\cos^2\theta^*\ + \cdots\,,
\ee
where $a$ is a positive $v$-dependent coefficient. Then, near $\cos\theta^* =
0$, i.e., the $\dX$ threshold,
\be\label{eq:dcstardX}
\frac{d\sigma}{d(\dX)} = \frac{\sigma}{2}\, \frac{d(\cos\theta^*)}{d(\dX)}
\propto \frac{1}{\sqrt{\dX - \dXm}}\,. 
\ee
It is the simple one-variable Taylor expansion of $\dX$ in
Eq.~(\ref{eq:dXexpand}) that has caused this singularity.

The discussion of $d\sigma/d(\dR)$ for the LSTC signal shares some features
with $d\sigma/d(\dX)$, though it it is qualitatively different. The $\dR$
distribution also vanishes below a threshold, $\dRm$, which is equal to $\dXm
= 2\cos^{-1}(v)$. This remarkable feature, derived in the appendix, can be
understood simply as a consequence of the fact that the minimum of $\dR$
occurs when {\em both} jet rapidities vanish. In that case, $\dR = \dphi =
\dX$.

At threshold, however, the $\dR$ distribution is $\propto \sqrt{\dR - \dXm}$,
not the inverse square root. As illustrated in Fig.~\ref{fig:dchidR}, it
rises sharply from threshold into a broad feature before decreasing. The
measure of the $\tpi$ velocity $v$ is given by the onset of the rise, not its
peak. This is the behavior seen in the CDF data in Fig.~\ref{fig:CDFdRjj},
where the rise starts very near $2\cos^{-1}(v) = 2.23$ for our input
masses. Both the $\dX$ and $\dR$ distributions measure the $\tpi$ velocity
$v$ and, therefore, provide confirmations of the $\tro \ra W\tpi$ hypothesis
which are {\em independent} of the background under the $\MWjj$ resonant peak
and of uncertainty in the $\etmiss$ resolution as well.

The reason for this qualitative difference between the two distributions is
that $d\sigma/\dX$ involves a one-dimensional trade of $\cos \theta^*$
for $\dX$, whereas $\dR$ is parametrized in terms of the three angles
$\theta, \theta^*,\phi^*$ in an intricate way, with all three being
integrated over to account for the constraint defining $\dR$.  In contrast to
what happens in the $\dX$ case, the Jacobian singularity at the threshold is
``antidifferentiated'' twice, hence its comparatively lower strength. Using a
Fadeev-Popov-like trick, the $\dR$ distribution can be written
\be\label{eq:dsigdR}
\frac{d \sigma}{d(\dR)} =    
\int d(\cos\theta) \, d(\cos\theta^*) d(\cos\phi^*) \,
\frac{d\sigma}{d(\cos\theta^*)} \, 
\delta\left(\dR - f(\cos\theta,\cos\theta^*,\cos\phi^*) \right)\,.
\ee
The function $f(\cos\theta,\cos\theta^*,\cos\phi^*)$ is shown in the
appendix to have its absolute minimum at $\cos\theta = \cos\theta^* =
\cos\phi^* = 0$, for which its value is equal to $\dXm$. Near its minimum it
is locally parabolic and its Taylor expansion is
\be\label{eq:dRexpand}
f(\cos\theta,\cos\theta^*,\cos\phi^*) = 
\dXm + {\thalf}\left(\bth\cos^2\theta + \bthst\cos^2\theta^* +
  \bphst\cos^2\phi^*\right) + \cdots 
\ee
The positive $v$-dependent coefficients $b_{\theta}, b_{\theta^*}$ and
$b_{\phi^{ *}}$ are also given in the appendix, Eq.~(\ref{eq:bterms}). For
$\dR$ close to $\dXm$, this expansion can be used to approximate
Eq.~(\ref{eq:dsigdR}). In a similar way as for the $\dX$ distribution,
integrating first over $\cos\theta^*$ generates the appearance of a Jacobian
inverse square root singularity $\propto [2(\dR - \dXm) - (\bth\cos^2\theta +
\bphst\cos^2\phi^*)]^{-1/2}$. The two remaining integrations over
$\cos\theta$ and $\cos\phi^*$ were trivial in the $\dX$ case as the integrand
did not depend on them, but this is not so for $\Delta R$ which involves a
double integration over a restricted angular phase space defined by
\be\label{eq:omega}
 0 \leq \bth\cos^2\theta + \bphst \cos^2\phi^* \leq 2 \left(\dR - \dXm
 \right)\,.
\ee
Performing the integral in Eq.~(\ref{eq:dsigdR}) near $\dRm = \dXm$ yields a
result $\propto \sqrt{\dR - \dXm}$.

We have examined the effect of finite jet masses (as opposed to jet
reconstruction and energy resolution) on the threshold values of the $\dR$
and $\dX$ distributions and the extraction of the $\tpi$ velocity $v$ from
them. Our jets (which include $b$-jets in the Pythia default $\tpi$-decay
table) have masses $\simle 10\,\gev$.  Assuming, for simplicity, equal jet
masses and denoting by $u = \sqrt{1-4M_{\rm jet}^2/M_{\tpi}^2}$ the jet
velocity in the $\tpi$ rest frame, the corrected $\dXm(u)$ is
\be\label{eq:dXcorr}
\dXm(u) = \cos^{-1}\frac{v^2 - u^2(1-v^2)} {v^2 + u^2(1-v^2)} \simeq
\cos^{-1}(2v^2 - 1) -v(1-v^2)^{1/2} (1-u^2)\,.
\ee
This is less than the massless $\dXm$ by half a percent for $M_{\rm jet} =
10\,\gev$.

Finally, as noted, the $\ta$ accounts for about 25--30\% of $W\tpi$
production. This decay gives a $\tpi$ velocity of 0.54 in the $\ta$ rest
frame and $\dXm = 2.00$. The effect is clearly visible in the $\dX$ and $\dR$
distributions for the primary parton in Fig.~\ref{fig:dchidR}, but is washed
out by the low-end tails for the reconstructed jets. We believe that the low
and high-end tails are due to the two $\tpi$ jets fragmenting to three jets
and the two leading jets being closer or farther apart than the original
pair. It turns out that our $Q$-value cut for $Z\tpi$ in Sec.~4 eliminates
the $a_T$ contribution to the signal.

\section*{3. The $\tro,\ta \ra W\tpi$ mode at the LHC}

As a reminder, we assumed $M_{\tro} = 290\,\gev$, $M_{\ta} = 1.1 M_{\tro} =
320\,\gev$, $M_{\tpi} = 160\,\gev$ and $\sin\chi = 1/3$ to describe the CDF
dijet excess. The Tevatron cross section is $2.2\,\pb$. At the 8~TeV LHC,
these parameters give $\sigma(W\tpi) = 9.5\,\pb$ ($2.0\,\pb$ for $W\ra e\nu,
\,\mu\nu$). These cross sections are 20\% higher than at 7~TeV, but this does
not translate into a 20\% increase in $S/B$. About 70\% of the LHC rate is
due to the $\tro$; the $\tro$ and $\ta$ interference is very small. For such
close masses, it is impossible to resolve the two resonances in the $\MWjj$
spectrum.

\begin{figure}[!t]
 \begin{center}
\includegraphics[width=3.15in, height=3.15in]{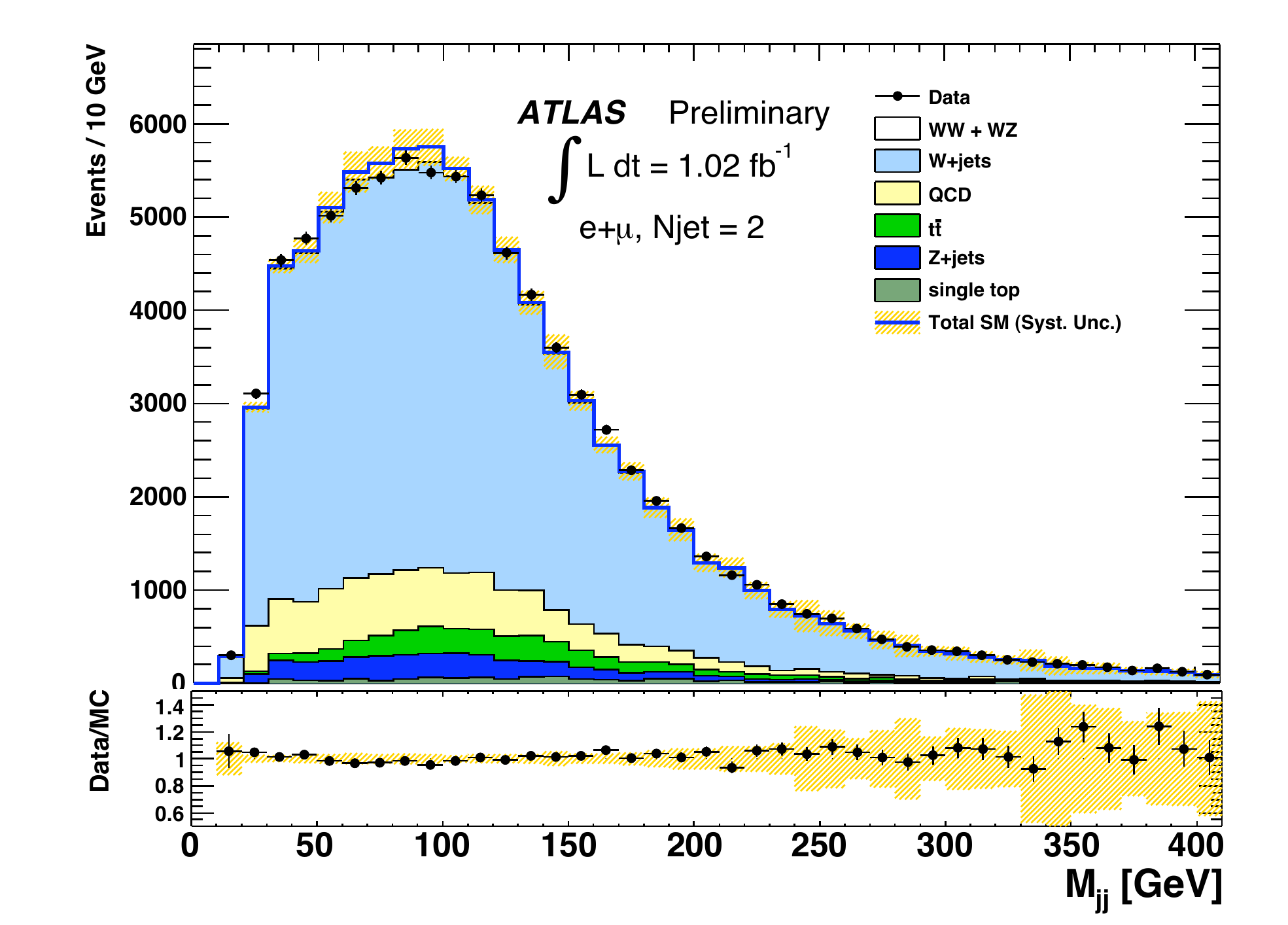}
\includegraphics[width=3.15in, height=3.15in]{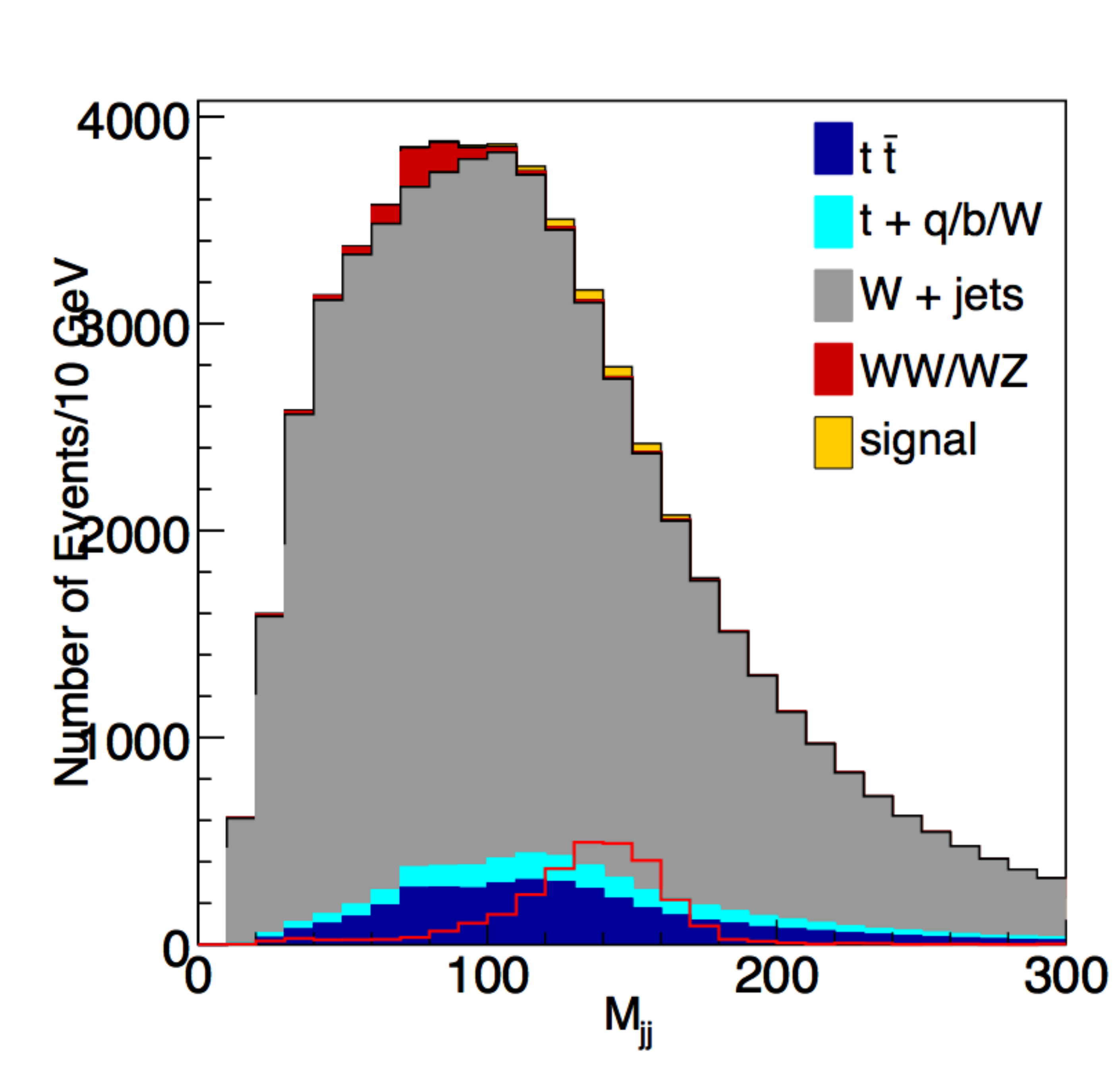}
\caption{Left: The ATLAS $\Mjj$ distribution for exactly two jets in $Wjj$
  production at $\sqrt{s} = 7\,\tev$ and $\int \CL dt = 1.02\,\ifb$; from
  Ref.~\cite{ATLASWjj} Right: Simulation of the $\Mjj$ distribution in $Wjj$
  production with ATLAS cuts (except that $p_T(\ell) > 30\,\gev$) for
  $1.0\,\ifb$. The open red histogram is the $\tpi\ra jj$ signal {\em times
    10.}}
\label{fig:LHCWjj}
 \end{center}
 \end{figure}

 Last summer, the ATLAS Collaboration published dijet spectra for
 $1.02\,\ifb$ of $Wjj$ data with exactly two jets and with two or more jets
 passing selection criteria~\cite{ATLASWjj}. The ATLAS cuts, taken as close
 to CDF's as practical, were: one isolated electron with $E_T > 25\,\gev$ or
 muon with $p_T > 20\,\gev$ and rapidity $|\eta_\ell| < 2.5$; $\etmiss >
 25\,\gev$ and $M_T(W) > 40\,\gev$; two (or more) jets with $p_T > 30\,\gev$
 and $|\eta_j| < 2.8$; and $p_T(jj) > 40\,\gev$ and $\deta < 2.5$ for the two
 leading jets. The $\Mjj$ distribution for the two-jet data is shown in
 Fig.~\ref{fig:LHCWjj}. There is no evidence of CDF's dijet excess near
 $150\,\gev$ nor even of the standard model $WW/WZ$ signal near
 $80\,\gev$. This is what we anticipated in Ref.~\cite{Eichten:2011xd}
 because of the great increase in $Wjj$ backgrounds at the LHC relative to
 the Tevatron. On the other hand, it is noteworthy and encouraging for future
 prospects that the ATLAS background simulation appears to fit the data well.

 In Fig.~\ref{fig:LHCWjj} we also show our simulation of the LSTC $\Mjj$
 signal and backgrounds at the LHC for $\sqrt{s} = 7\,\tev$ and $\int \CL dt
 = 1.0\,\ifb$.  ATLAS's cuts were used except that we required $p_T(\ell) >
 30\,\gev$.\footnote{Backgrounds were generated at matrix-element level using
   ALPGENv213~\cite{Mangano:2002ea}, then passed to {\sc Pythia}v6.4 for
   showering and hadronization. We use CTEQ6L1 parton distribution functions
   and a factorization/renormalization scale of $\mu = 2 M_W$ throughout. For
   the dominant $W+$jets background we generate $W+2j$ (exclusive) plus
   $W+3j$ (inclusive) samples, matched using the MLM procedure~\cite{MLM}
   (parton level cuts are imposed to ensure that $W+0, 1$ jet events cannot
   contribute).  After matching, the overall normalization is scaled to the
   NLO $W+jj$ value, calculated with MCFMv6~\cite{Campbell:2011bn}. After
   passing through {\sc Pythia}, final state particles are combined into
   $(\eta, \phi)$ cells of size $0.1\times 0.1$, and the energy in each cell
   smeared with $\Delta E/E = 1.0/\sqrt{E/{\gev}}$. The energy of each cell
   is rescaled to make it massless. Isolated photons and leptons ($e,\mu$)
   are removed, and all remaining cells with energy greater than $1\,\gev$
   are clustered into jets using FastJet (anti-kT algorithm, $R =
   0.4$)~\cite{Cacciari:2005hq}. Estimates of the background including higher
   order effects have been shown to be completely consistent with our LO+PS
   treatment~\cite{Campbell:2011gp,Frederix:2011ig}. Finally, the quadratic
   ambiguity in the $W$ reconstruction was resolved by choosing the solution
   with the smaller $p_z(\nu)$.} This tighter cut and our inability to
 include the data-driven QCD background account for our lower event rate
 compared to ATLAS. Despite this, the agreement between the two is quite
 good. In particular, our simulation shows that the CDF/ATLAS cuts can
 neither reveal nor exclude the LSTC interpretation of the CDF signal at the
 LHC for any reasonable luminosity.\footnote{Models of the CDF signal that
   are $gg$-initiated or involve large coupling to heavier quarks, e.g.,
   Refs.~\cite{Gunion:2011bx, Nelson:2011us}, are likely excluded by the
   ATLAS data.}

\begin{figure}[!t]
 \begin{center}
\includegraphics[width=3.15in, height=3.15in]{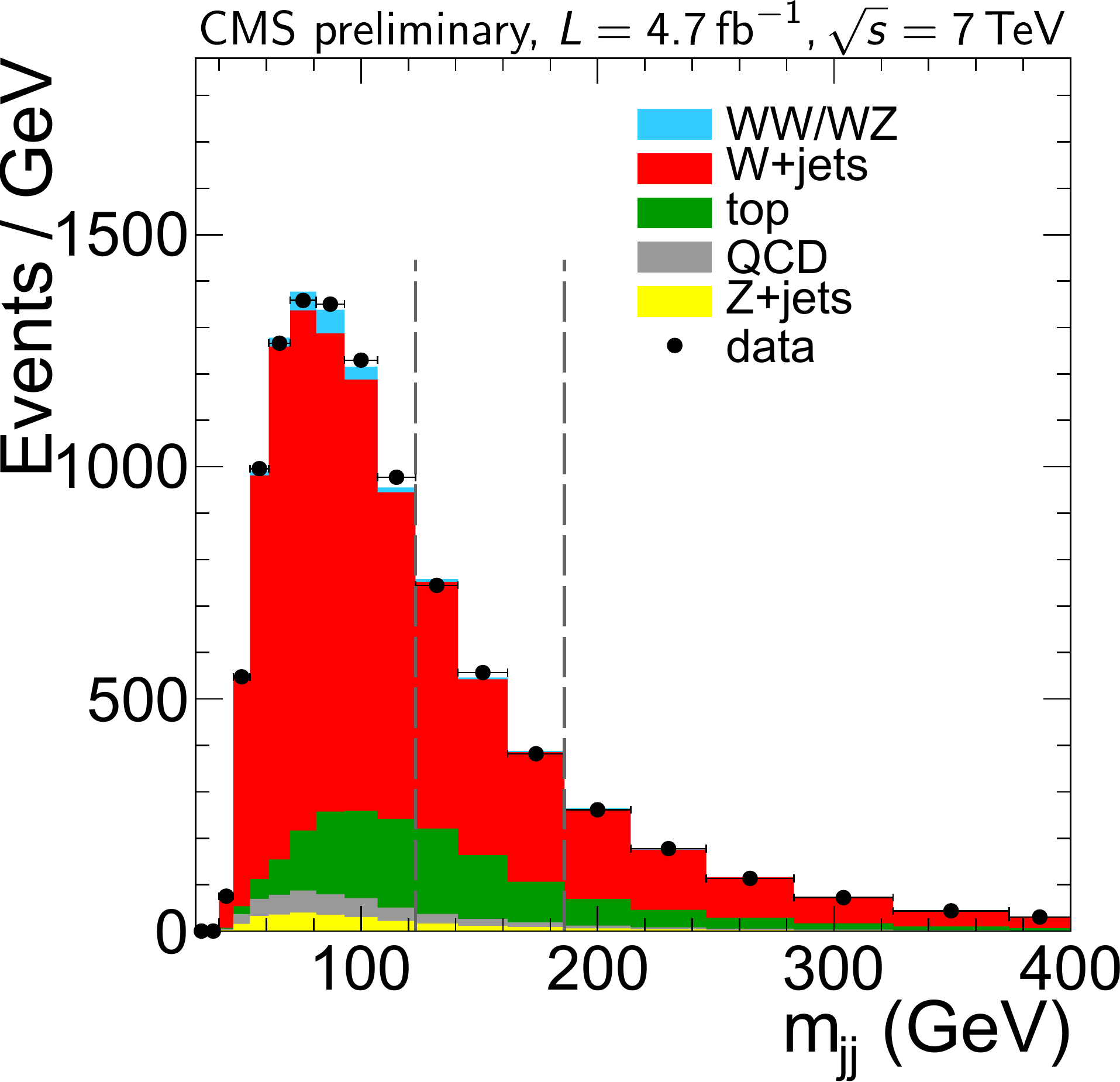}
\includegraphics[width=3.15in, height=3.15in]{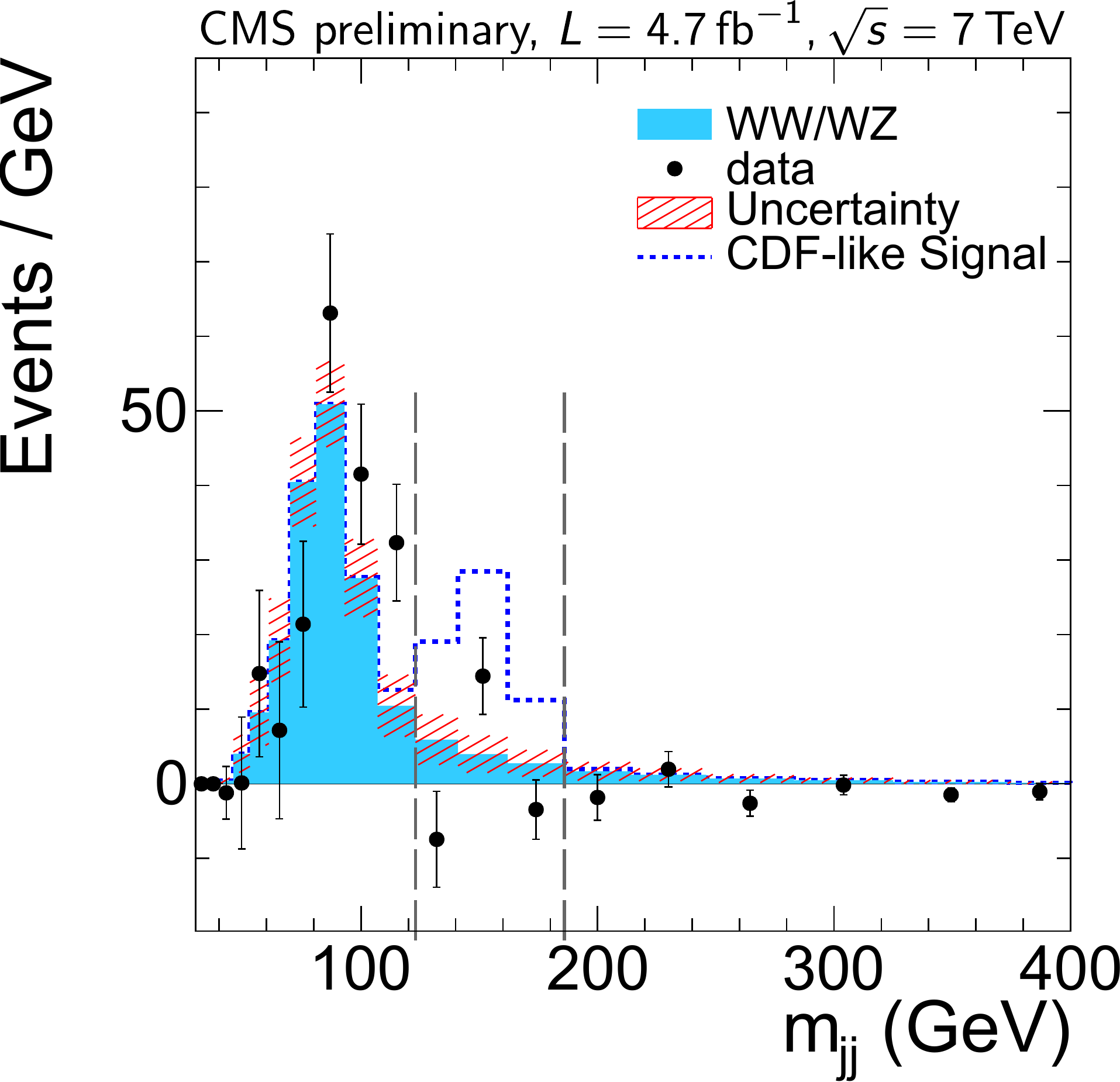}
\includegraphics[width=3.15in, height=3.15in]{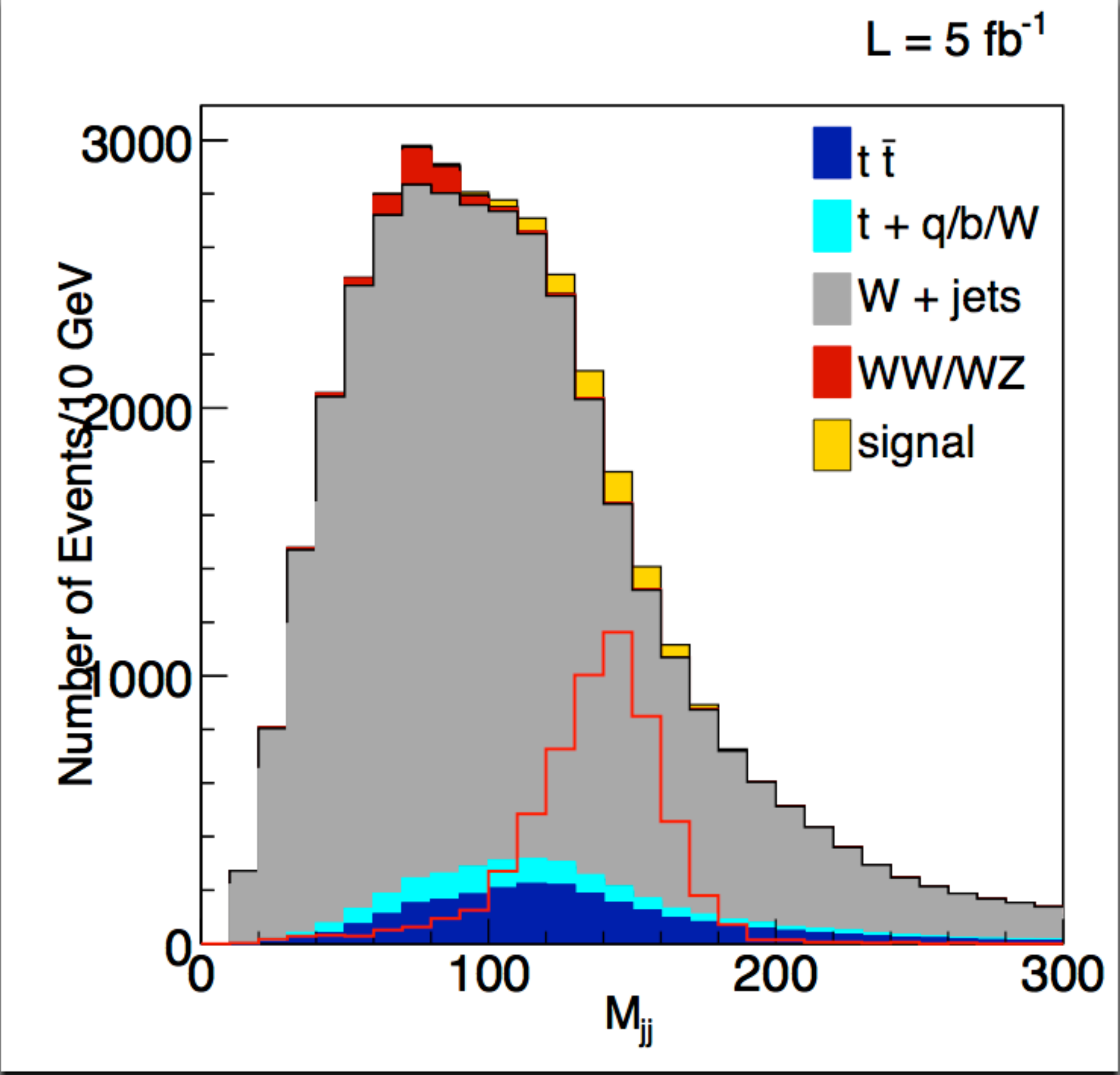}
\caption{The CMS $\Mjj$ distributions for $4.7\,\ifb$ of $W\to \mu\nu,\,
  e\nu$ plus two or three jets data at $\sqrt{s} = 7\,\tev$ before (top left)
  and after (top right) the background subtraction summarized in the text;
  from Ref.~\cite{CMSWjj}. On the bottom is our $\Mjj$ distribution for the
  $\tro,\ta \ra W\tpi \ra \ell\nu_\ell jj$ signal and backgrounds at the LHC
  for $5\,\ifb$. Augmented ATLAS-like cuts as described in the text were
  used. The open red histograms are the $\tpi$ and $\tro$ signals {\em times
    10}.
  \label{fig:CMSELMPWjj}}
 \end{center}
 \end{figure}

 Recently, the CMS Collaboration studied the dijet-mass spectrum in $W(\to
 \ell\nu)$ plus jets production with $4.7\,\ifb$ at
 $7\,\tev$~\cite{CMSWjj}. CMS used the following cuts which were partly
 adopted from Ref.~\cite{Eichten:2012br}: $p_T(e,\mu) > 25, 30\,\gev$ and
 rapidity $|\eta(e,\mu)|< 2.5, 2.1$, $\Delta R(\ell,j) > 0.3$;
 $\etmiss(e,\mu) > 35, 25\,\gev$, $\Delta\phi(\etmiss,j) > 0.4$; $M_T(W) >
 50\,\gev$ and $p_T(W) > 60\,\gev$; exactly two or three jets with $p_{T1} >
 40\,\gev$, $p_{T2,3} > 30\,\gev$, $|\eta_j| < 2.4$; and $p_T(jj) >
 45\,\gev$, $\Delta\eta(jj) < 1.2$. CMS used {\sc MadGraph} to generate
 $W+{\rm jets}$ and a data-driven method to determine the $\Mjj$ shape and
 background: A superposition of a set of templates was constructed in which
 the {\sc MadGraph} factorization and renormalization scales were varied up
 and down by a factor of two from their default values, and this was fit to
 the dijet spectrum {\em outside} the signal region, taken to be 123 to
 $186\,\gev$. The $Wjj$ background in the signal region was then determined
 from this fit. The CMS dijet spectra before and after background subtraction
 are shown in Fig.~\ref{fig:CMSELMPWjj}. Note that the vertical scale is
 ``Events/GeV.''  No significant enhancement near $150\,\gev$ was
 observed. (What CMS meant by a ``CDF-like signal'' is not specified in
 Ref.~\cite{CMSWjj}.) Using a $WH$ production model, CMS reported a 95\%
 upper limit on the production cross section times $B(W \to \ell\nu$) of
 $1.3\,\pb$.

 We studied the LSTC $Wjj$ signal at $\sqrt{s} = 7\,\tev$ in
 Ref.~\cite{Eichten:2012br}, before the CMS paper's release. Our prediction
 for the cross section was $\sigma B = 1.7\,\pb$, 30\% higher than CMS's
 limit. In order to achieve a better outcome than ATLAS's 2011 study, we
 examined a variety of cuts motivated by $\tro \ra W\tpi$ kinematics. Cuts
 quite similar to those we proposed for the Tevatron in
 Ref.~\cite{Eichten:2011sh} typically caused the background to peak very near
 the dijet resonance. To get the signal off the peak (and more like the
 original CDF $\Mjj$ excess~\cite{Aaltonen:2011mk}), we used the following:
 lepton $p_{T\ell} > 30\,\gev$ and $|\eta_\ell|< 2.5$, $\etmiss > 25\,\gev$,
 $M_T(W) > 40\,\gev$ and $p_T(W) > 60\,\gev$; exactly two jets with $p_{T1} >
 40\,\gev$, $p_{T2} > 30\,\gev$, $|\eta_j| < 2.8$; $p_T(jj) > 45\,\gev$,
 $\Delta\eta(jj) < 1.2$; and $Q = \MWjj - \Mjj - M_W < 100\,\gev$. The
 resulting $\Mjj$ distribution is also displayed in
 Fig.~\ref{fig:CMSELMPWjj}. Counting events in the range $120 < \Mjj <
 170\,\gev$ gives $S/\sqrt{B} = 6.5$ for this luminosity, but only $S/B =
 0.050$. The $\dR$ and $\dX$ signals are also small and not useful. Because
 of the small $S/B$, and in view of the difficulty CMS had fitting the dijet
 spectrum in the diboson and CDF-signal region, we believe that a better
 understanding of the backgrounds is required to observe or exclude the LSTC
 signal in this channel. 

%
\begin{figure}[!t]
 \begin{center}
\includegraphics[width=3.15in, height=3.15in]{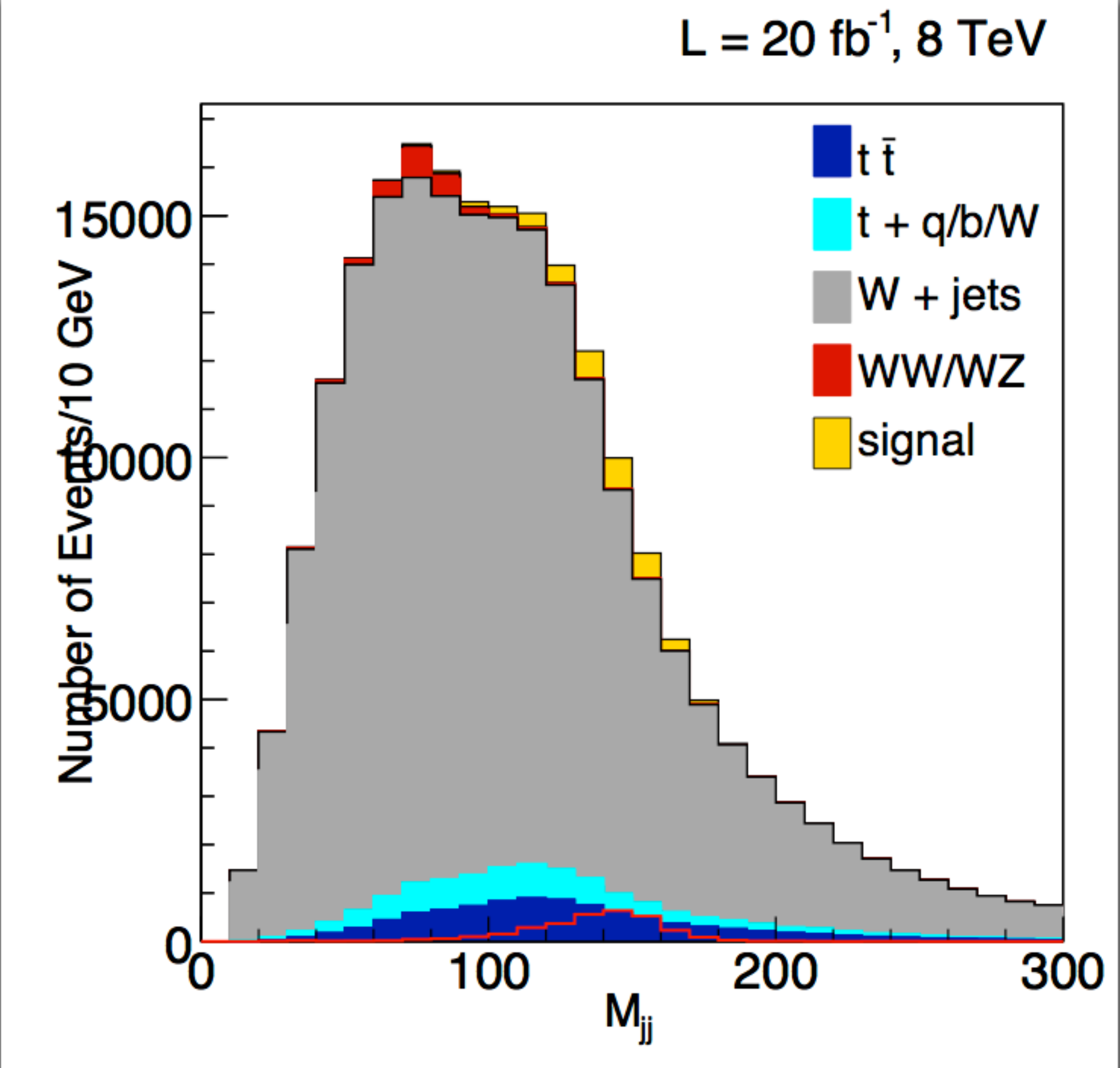}
\includegraphics[width=3.15in, height=3.15in]{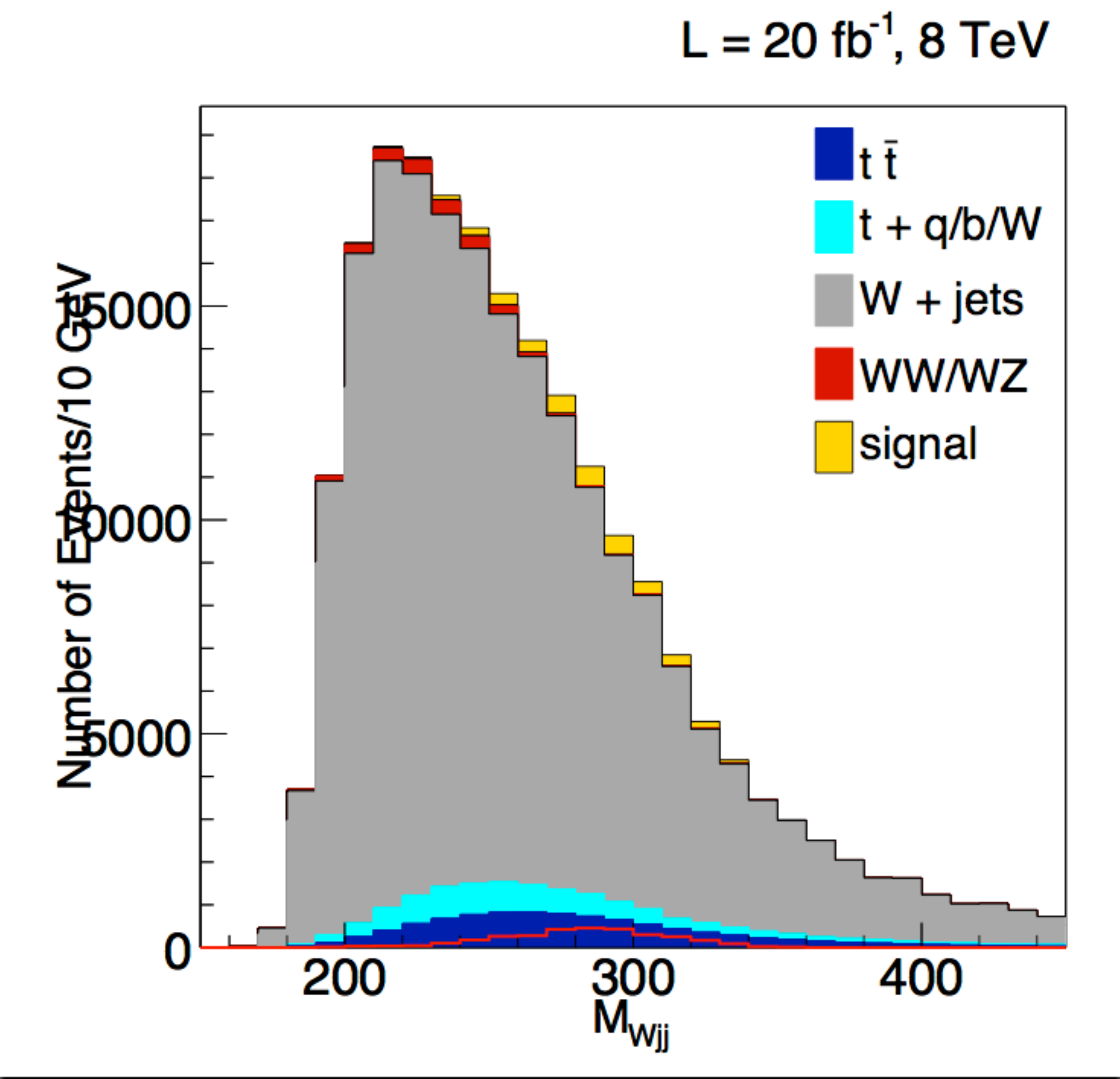}
\caption{The $\Mjj$ and $\MWjj$ distributions of $\tro,\ta \ra W\tpi \ra
  \ell\nu_\ell jj$ and backgrounds at the LHC for $\sqrt{s} = 8\,\tev$ and
  $\int \CL dt = 20\,\ifb$. Augmented ATLAS-like cuts as described in the
  text are employed. The open red histograms are the unscaled $\tpi$ and
  $\tro$ signals.
  \label{fig:ELMPWjj}}
 \end{center}
 \end{figure}
%


 Our simulations of the $\Mjj$ and $\MWjj$ distributions in $Wjj$ production
 at $\sqrt{s} = 8\,\tev$ are shown in Fig.~\ref{fig:ELMPWjj} for $\int \CL dt
 = 20\,\ifb$. The same cuts as above are used. Counting events in the range
 $120 < \Mjj < 170\,\gev$ gives $S/\sqrt{B} = 10.2$ for this luminosity but
 still only $S/B = 0.050$. Despite this large ``significance'', we remain
 uncertain of the ability of the $\ell\nu jj$ channel to settle the questions
 of CDF's dijet excess and our interpretation of it.

\section*{4. The $\tropm,\tapm \ra Z \tpipm$ mode}

In view of this situation with the $W\tpi$ signal, observation of the isospin
partner $\tropm,\tapm \ra Z\tpipm$ of the $W\tpiz$ mode can provide the
needed test of the LSTC interpretation of CDF's $Wjj$ signal. At the LHC, we
predict $\sigma(\tropm,\tapm \ra Z\tpipm) = 2.8\,\pb$, lower than
$\sigma(\tropm,\tapm \ra W\tpiz) = 4.1\,\pb$ because of the reduced phase
space, $\propto p^3$. Then, $\sigma(\tropm,\tapm \ra Z\tpi \ra \ellp\ellm jj)
= 190\,\fb$ for $\ell = e$ and $\mu$, of which, 80\% is due to the
$\tropm$. This rate is about 10\% of the $W\tpi \ra \ell\nu_\ell jj$
signal. We might expect, therefore, that $\sim 10$ times the luminosity
needed for the $W\tpi$ signal would be required for the same sensitivity to
$Z\tpi$. Actually, the situation is better than this because there is no QCD
multijet background nor $\etmiss$ resolution to pollute the $Zjj$ data.

\begin{figure}[!t]
 \begin{center}
\includegraphics[width=3.15in, height=3.15in]{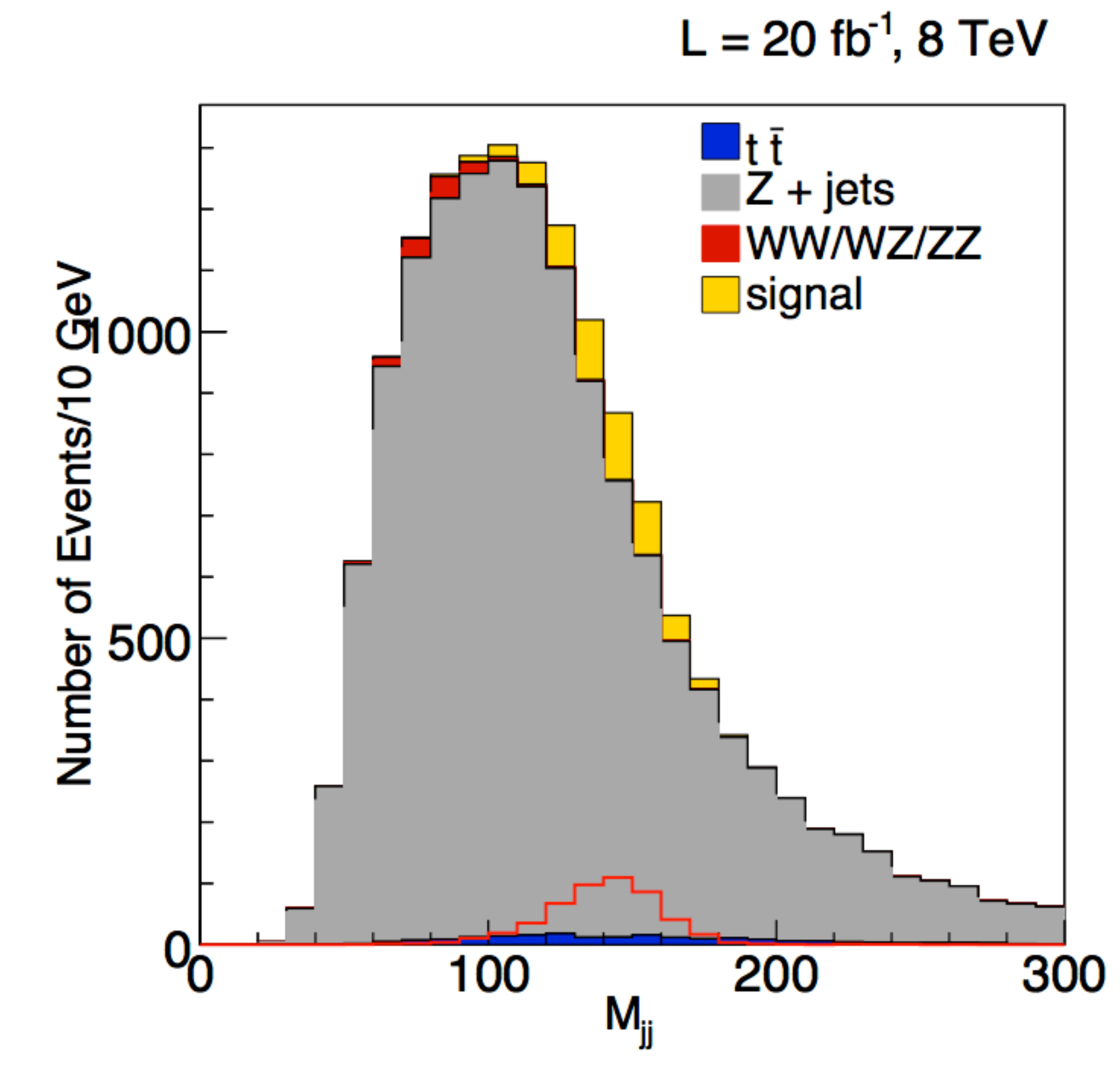}
\includegraphics[width=3.15in, height=3.15in]{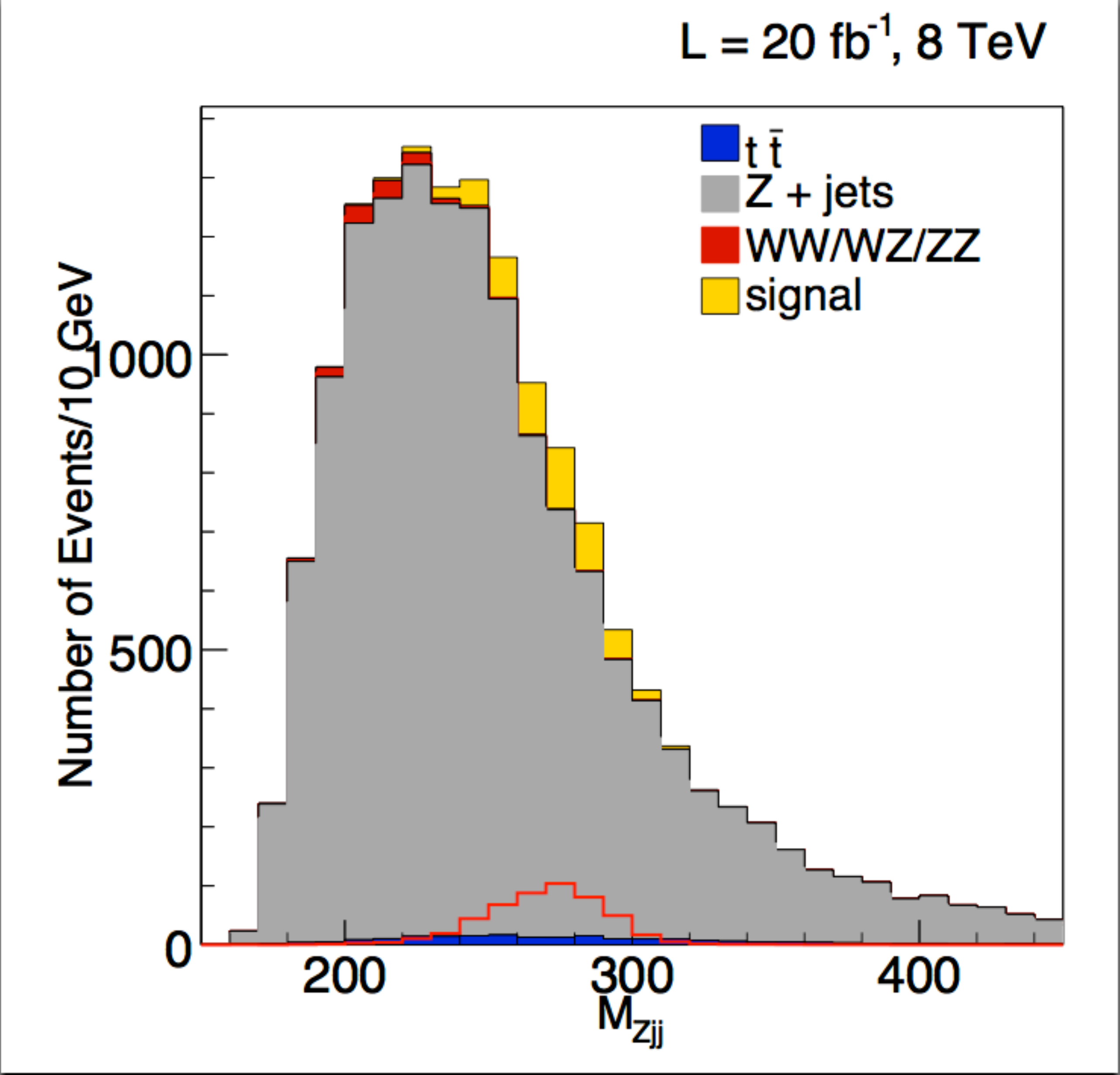}
\caption{The $\Mjj$ and $\MZjj$ distributions of $\tropm \ra Z\tpipm \ra
  \ellp\ellm jj$ and backgrounds at the LHC for $\sqrt{s} = 8\,\tev$ and
  $\int \CL dt = 20\,\ifb$. The cuts used are described in the text. The open
  red histograms are the $\tpi$ and $\tro$ signals.
  \label{fig:ELMPZjj}}
 \end{center}
 \end{figure}
\begin{figure}[!ht]
 \begin{center}
\includegraphics[width=3.15in, height=3.15in]{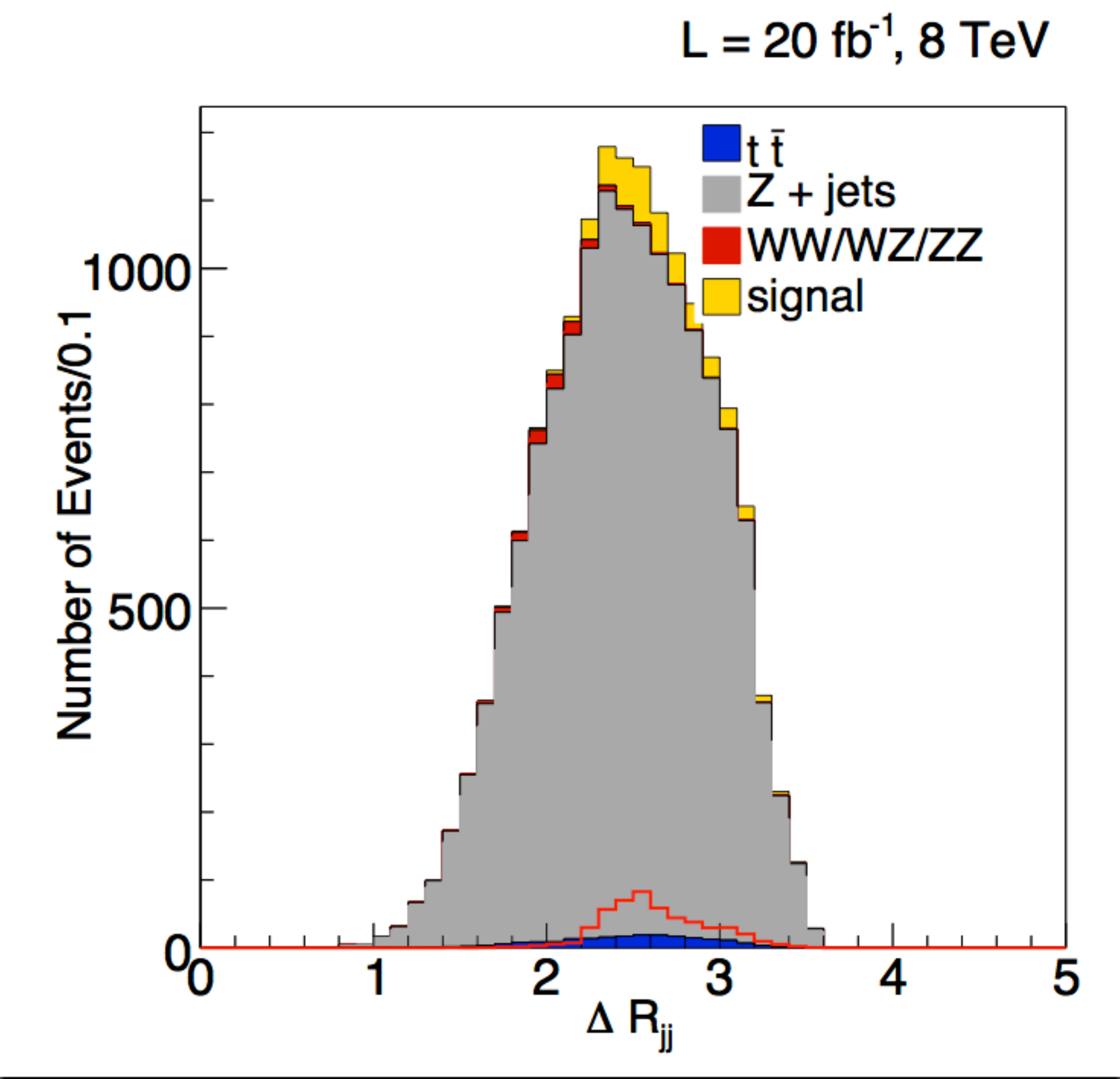}
\includegraphics[width=3.15in, height=3.15in]{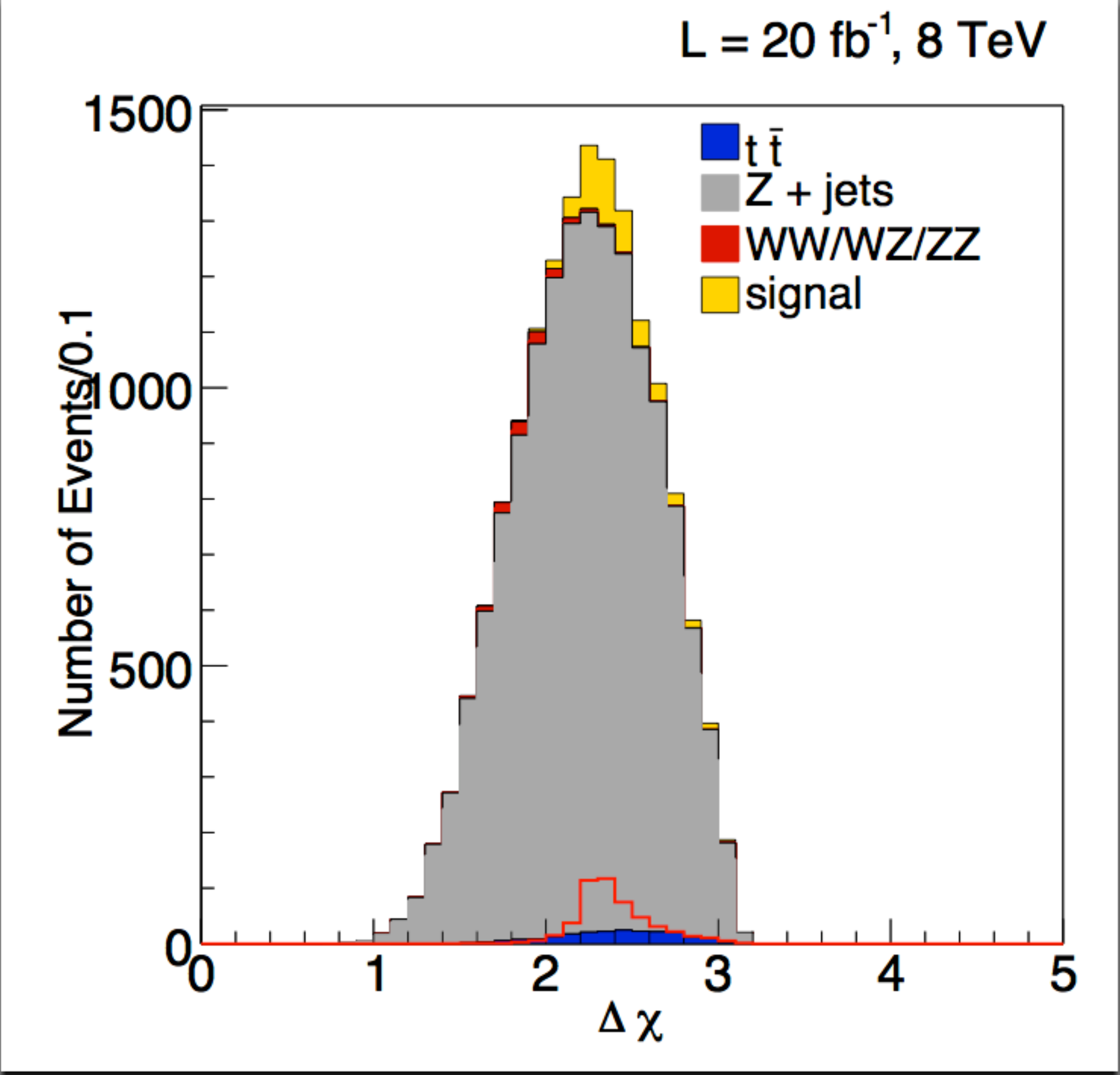}
\caption{The $\dR$ and $\dX$ distributions for $\tropm \ra Z\tpipm \ra
  \ellp\ellm jj$ and backgrounds at the LHC for $\sqrt{s} = 8\,\tev$ and
  $\int \CL dt = 20\,\ifb$. The cuts used are described in the text. The open
  red histograms are the signals.
  \label{fig:ELMPZjjRX}}
 \end{center}
 \end{figure}

 Figure~\ref{fig:ELMPZjj} shows the $Z\tpi$ signal and its background, almost
 entirely from $Z+{\jets}$, for $\sqrt{s} = 8\,\tev$ and $\int\CL dt =
 20\,\ifb$. The cuts used here are: two electrons or muons of opposite charge
 with $p_T > 30\,\gev$, $|\eta_\ell| < 2.5$, $80 < M_{\ellp\ellm} <
 100\,\gev$ and $p_T(Z) > 50\,\gev$; exactly two jets with $p_T > 30\,\gev$
 and $|\eta_j| < 2.8$; $p_T(jj) > 40\,\gev$, $\Delta\eta(jj) < 1.75$; and $Q =
 \MZjj - \Mjj - M_Z < 60\,\gev$. This $Q$-cut is very important in reducing
 the background. However, it excludes the 20\% of $Z\tpi$ that comes from
 $\tapm$ production.\footnote{We considered $Q < 80\,\gev$ to include the
   $\ta$, but found that the background increased substantially faster than
   the signal.  The $\tro,\ta \ra WZ \ra \ellp\ellm jj$ process is included
   in this simulation, but it also is removed by the $Q$-cut.} These give
 $S/\sqrt{B} = 6.2$ and $S/B = 0.11$ for the dijet signal in $120 < \Mjj <
 170\,\gev$. The figure also shows the $\MZjj$ distribution; it has
 $S/\sqrt{B} = 6.4$ and $S/B = 0.12$ for $250 < \MZjj < 320\,\gev$. These
 signal-to-background rates and the position of the dijet signal on the
 falling backgrounds are similar to those in Ref.~\cite{CDFnew}. Therefore,
 if our interpretation of the CDF dijet excess is correct, both $\tpi \to jj$
 and $\tro \to \ellp\ellm jj$ will be observable soon.

 Figure~\ref{fig:ELMPZjjRX} shows the $\dR$ and $\dX$ distributions for
 $\tro \ra Z\tpi \ra \ellp\ellm jj$. The skyscraper-shaped $\dX$
 distribution is especially interesting. The background peaks at $\dX \simeq
 2.3$, and appears rather symmetrical about this point except that its high
 side falls more rapidly above $2.7$ because $(\dX)_{\rm max} = \pi$. The
 signal's $\dX$ distribution sits atop the skyscraper, concentrated in about
 330~events in three bins at $\dX = 2.2$--2.4, whereas the theoretical $\dXm
 = 2\cos^{-1}(v) = 2.31$ for $\tro \to Z\tpi$. This is just as expected when
 jet reconstruction is taken into account; see Fig.~\ref{fig:dchidR}. If the
 actual $\dX$ data, with our cuts, has the shape of our simulation, we
 believe the signal excess can be observed. Similar remarks apply to the
 shape and observability of the slightly broader $\dR$ distribution in
 Fig.~\ref{fig:ELMPZjjRX}.

\section*{5. The $\tropm,\tapm \to WZ$ mode}

Finally, the decay channel $\tropm,\tapm \to W^\pm Z$ furnishes another
important check on the LSTC hypothesis provided that $\sin\chi \simge
1/4$. The dominant contribution, $\tro \to W_L Z_L$, has an angular
distribution $\propto \sin^2\theta$ so that the production is fairly
central. We expect $\sigma(\tro,\ta \to WZ)/\sigma(\tro,\ta \to W\tpiz)
\simeq (p(Z)/p(\tpi))^3\, \tan^2\chi$. The {\sc Pythia} rates are roughly
consistent with this. For our input masses and $\sin\chi =
({\tfifth},{\tfourth},{\tthird},{\thalf})$, we obtain the following cross
sections:
\bea\label{eq:WZrates}
\sigma(\tro,\ta \to WZ \to \ellp\ellm\ellpm\nu_\ell) &=&
(9,\,15,\,26,\,54)\,\fb\,, \\
\sigma(\tro,\ta \to WZ \to \ellp\ellm jj) &=& (27,\,48,\,80,\,170)\,\fb \,,\\
\sigma(\tro,\ta \to WZ \to \ell\nu jj) &=& (90,\,155,\,260,\,555)\,\fb \,,\\
\sigma(\tro,\ta \to WW \to \ell\nu jj) &=& (140,\,220,\,380,\,795)\,\fb \,,\\
\sigma(\tro,\ta \to Z\tpi \to \ellp\ellm jj) &=& (205,\,200,\,190,\,145)\,\fb \,,
\eea
for $\ell = e,\mu$.

The $\tro,\ta \to \ellp\ellm\ellpm\nu_\ell$ mode has been discussed in
Refs.~\cite{Brooijmans:2008se, Brooijmans:2010tn}. It has the advantages of
cleanliness and freedom from jet uncertainties (except $\etmiss$
resolution). Standard-model $WZ$ production at the LHC peaks at $M_{WZ} =
300\,\gev$~\cite{Eichten:1984eu}, near $M_{\tro}$, and this is the dominant
background to the $3\ell\nu$ signal. The D\O\ collaboration searched for this
channel using the standard LSTC parameters including $\sin\chi = 1/3$, and
excluded it at 95\% C.L.~up to $M_{\tro} \simeq 400\,\gev$ so long as the
$\tro \to W\tpi$ channel is closed~\cite{Abazov:2009eu}.

\begin{figure}[!t]
 \begin{center}
\includegraphics[width=3.15in, height=3.15in]{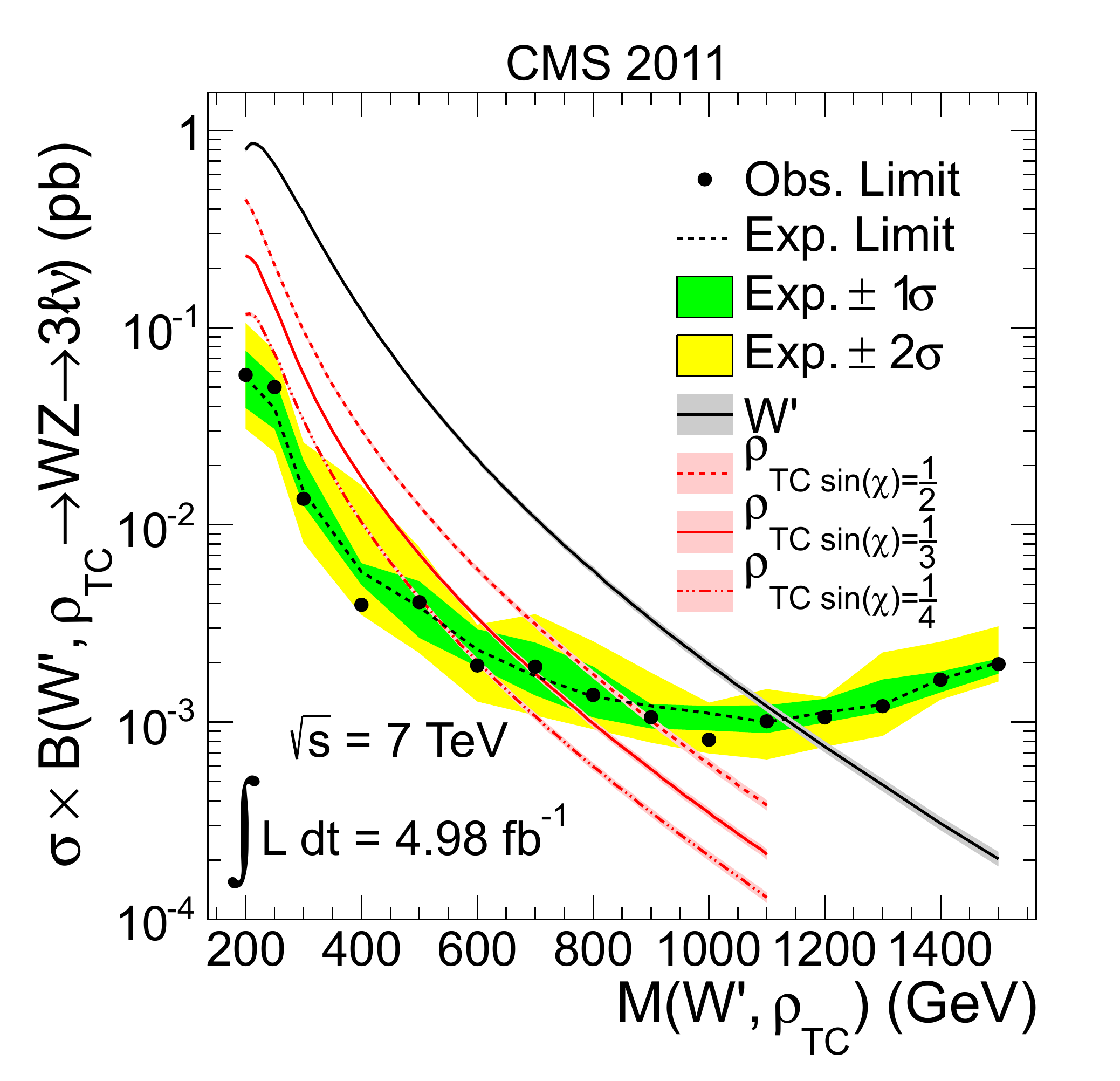}
\includegraphics[width=3.15in, height=3.15in]{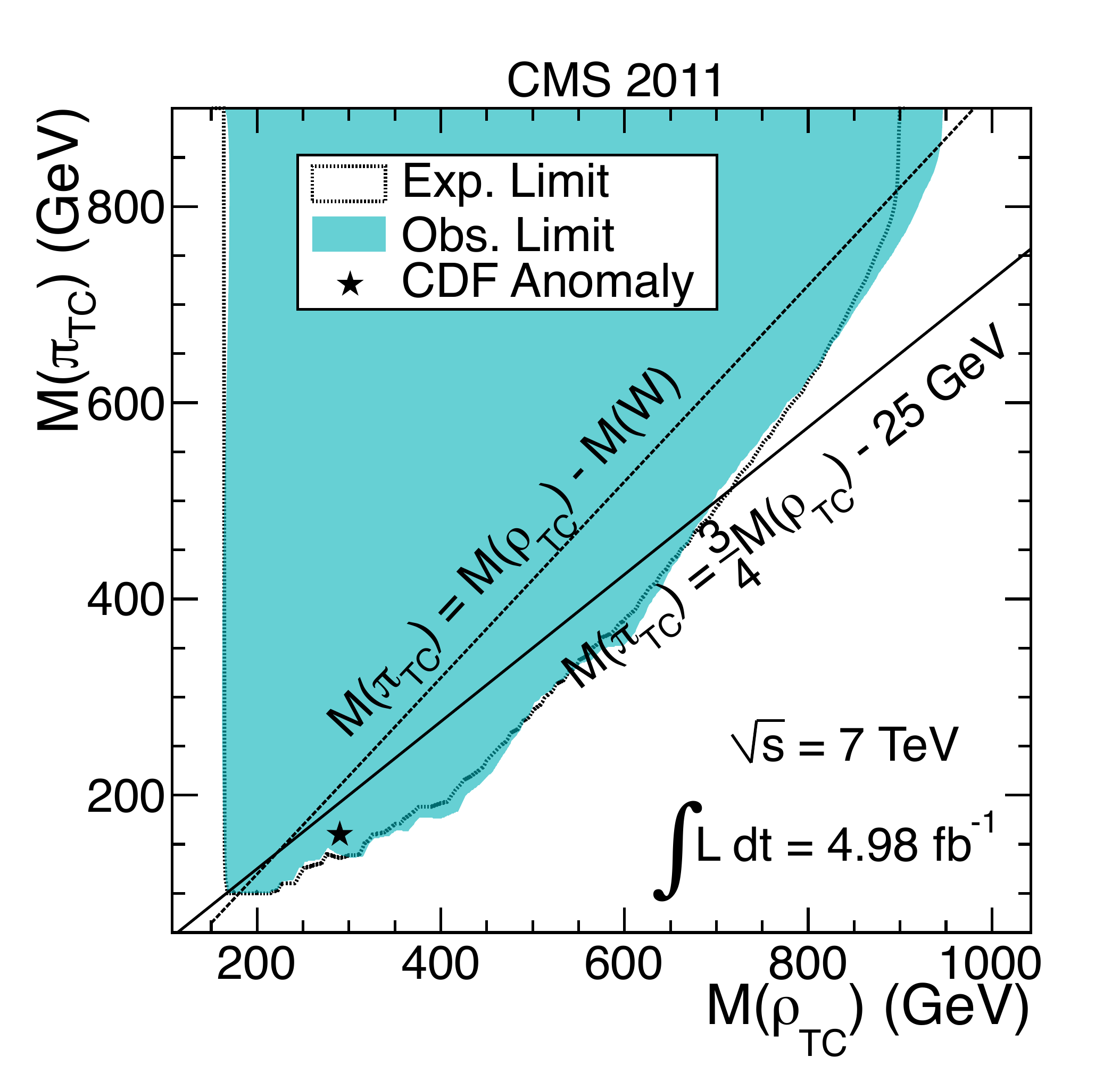}
\caption{Left: CMS $WZ \to 3\ell\nu$ cross section limits for $\int \CL dt =
  4.98\,\ifb$ at $\sqrt{s} = 7\,\tev$. The LSTC limit curves for $\sin\chi =
  {\tfourth},{\tthird},{\thalf}$ assume that $M_{\tpi} = 0.75 M_{\tro} -
  25\,\gev$. Right: Two-dimenional exclusion plot for LSTC with $\sin\chi =
  1/3$ as described in the text. The CDF mass point is marked by the
  star. From Ref.~\cite{Collaboration:2012kk}.}
  \label{fig:CMS}
 \end{center}
 \end{figure}

The CMS Collaboration recently reported a search for a sequential standard
model $W'$ and for $\tro,\ta \to WZ \to 3\ell\nu$ using $4.98\,\ifb$ of
$7\,\tev$ data~\cite{Collaboration:2012kk}. The cross section limits and
$M_{\tro}$ vs.~$M_{\tpi}$ exclusion plot are shown in Fig.~\ref{fig:CMS}. The
LSTC limit curves for $\sin\chi = {\tfourth},{\tthird},{\thalf}$ assume that
$M_{\tpi} = 0.75 M_{\tro} - 25\,\gev$. This stringent assumption
significantly enhances $B(\tropm \to WZ)$ above its value for the CDF mass
point. For the 2-D exclusion plot, standard LSTC parameters, including
$\sin\chi = 1/3$, were used. The CDF mass point is indicated by the star. We
predicted $21\,\fb$ for the signal at $7\,\tev$. Applying a $k$-factor of
1.36 in this mass range, CMS excludes $M_{\tpi} > 140\,\gev$ at the 95\%
C.L.~for $M_{\tro} = 275$--$290\,\gev$. The 95\% upper limit on the cross
section at $M_{\tro} = 290\,\gev$ is about $20\,\fb$. Using the CMS
$k$-factor, we estimate that the CDF point is allowed for $\sin\chi \simle
0.30$.

\begin{figure}[!t]
 \begin{center}
\includegraphics[width=3.15in, height=3.15in]{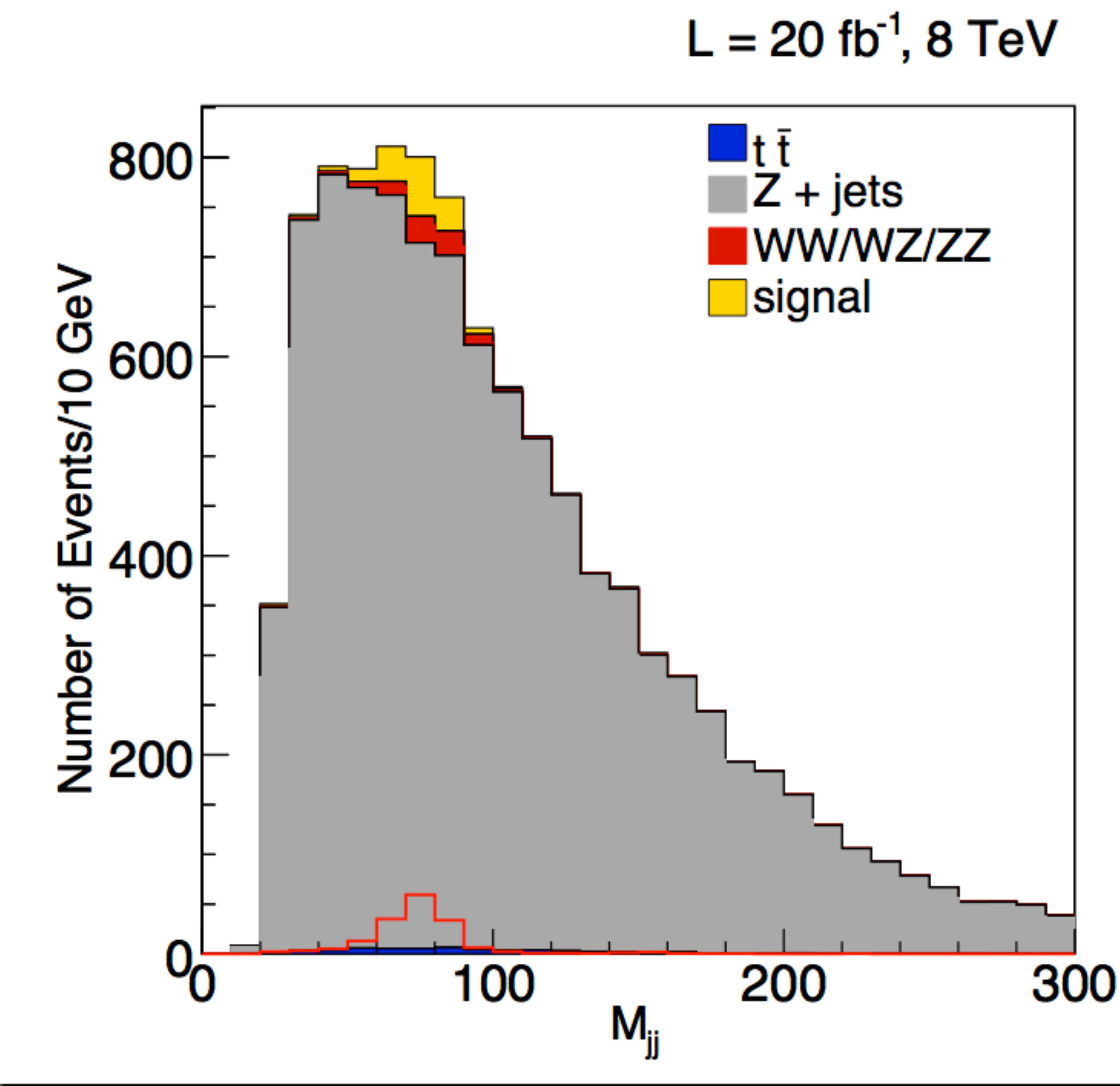}
\includegraphics[width=3.15in, height=3.15in]{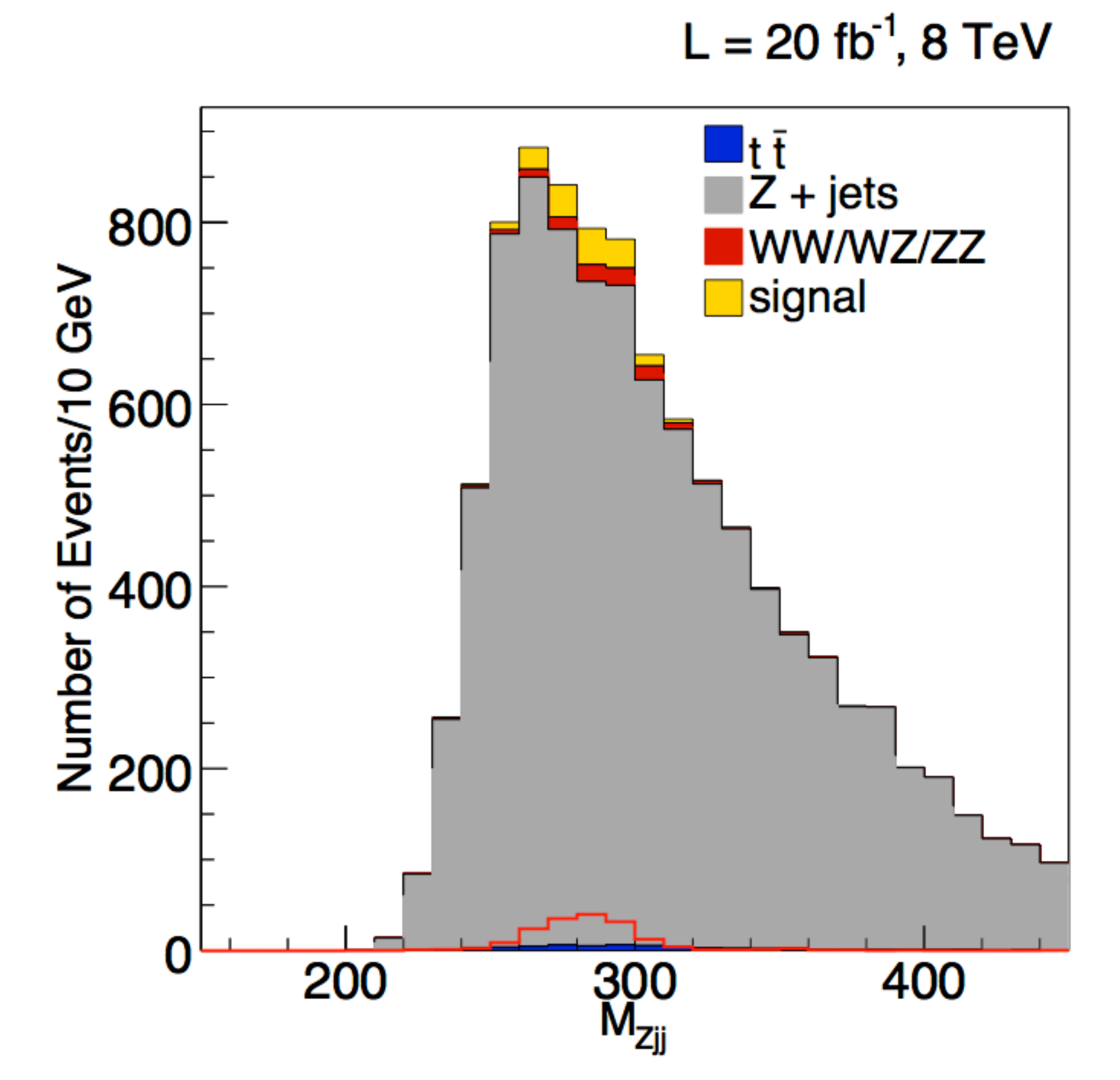}
\caption{The $\Mjj$ and $\MZjj$ distributions of $\tropm,\tapm \to WZ \to
  \ellp\ellm jj$ and backgrounds at the LHC for $\sqrt{s} = 8\,\tev$ and $\int
  \CL dt = 20\,\ifb$. The cuts used are described in the text. The open red
  histograms are the $\tpi$ and $\tro$ signals.
  \label{fig:WZMdists}}
 \end{center}
 \end{figure}
\begin{figure}[!t]
 \begin{center}
\includegraphics[width=3.15in, height=3.15in]{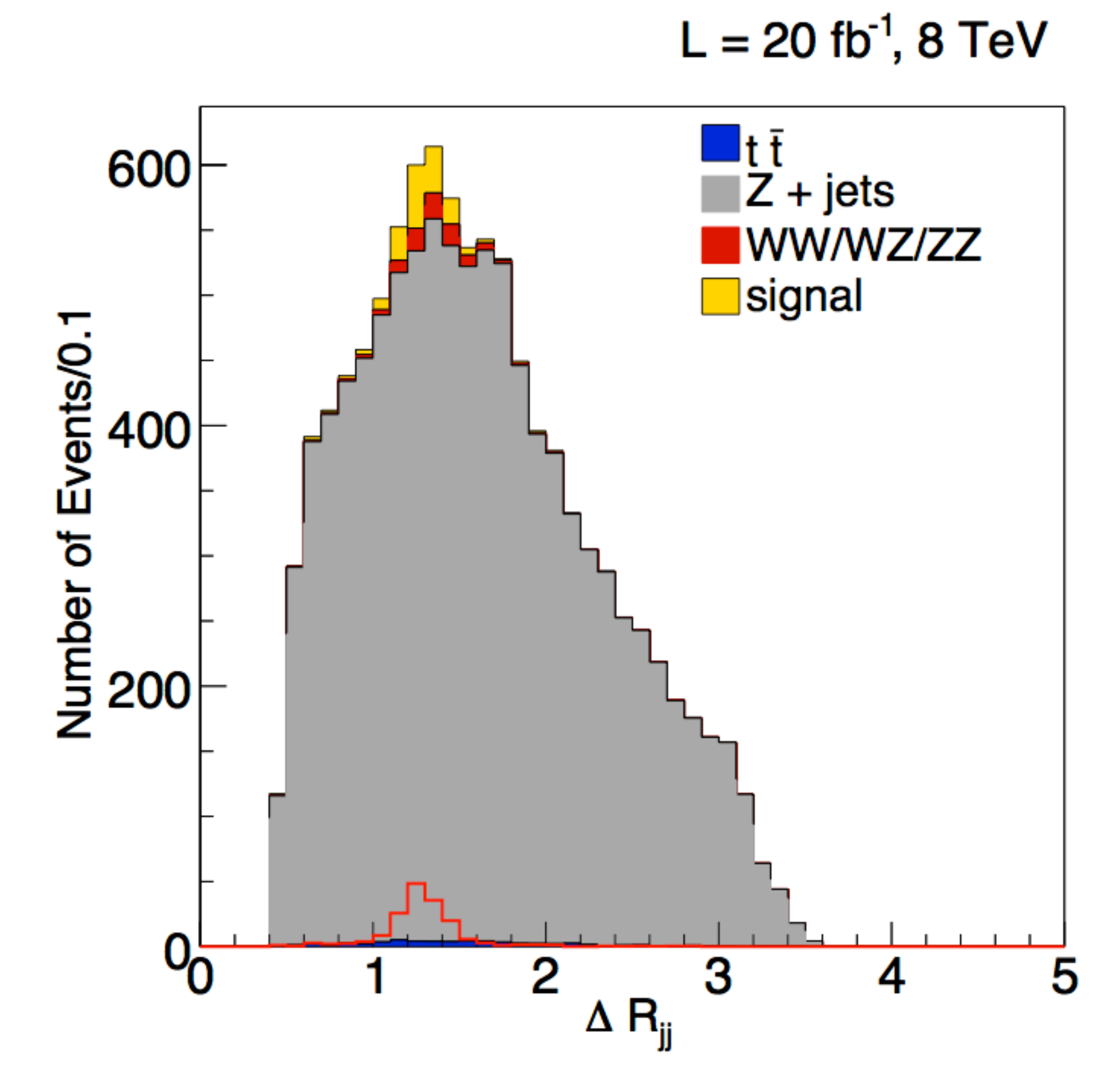}
\includegraphics[width=3.15in, height=3.15in]{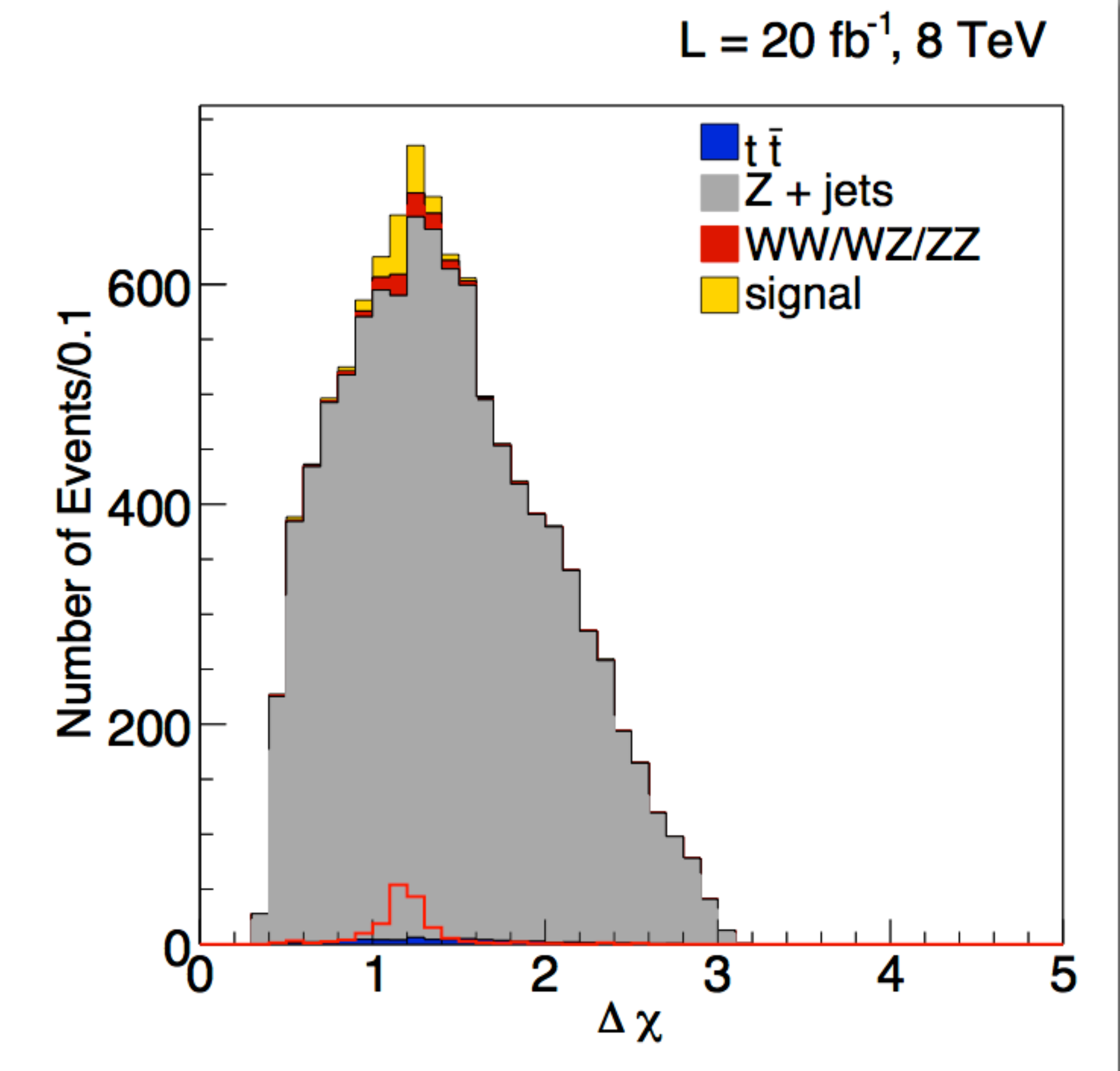}
\caption{The $\dR$ and $\dX$ distributions of $\tropm,\tapm \to WZ \to
  \ellp\ellm jj$ and backgrounds at the LHC for $\sqrt{s} = 8\,\tev$ and
  $\int \CL dt = 20\,\ifb$. The cuts used are described in the text. The open
  red histograms are the signals.
  \label{fig:WZRXdists}}
 \end{center}
 \end{figure}

 The dominant background to $\tro,\ta \to WZ \to \ellp\ellm jj$ is $Z +
 \jets$. As can be inferred from Fig.~\ref{fig:LHCWjj} for $Wjj$ production
 with ATLAS/CDF cuts, the signal will sit at the top of the $\Mjj$
 spectrum. This is what makes the dijet signal in $WW/WZ \to \ell\nu jj$ so
 difficult to see. On the plus side, since the LSTC and standard model
 diboson processes have very similar production characteristics, the two
 signals can be seen with the same cuts and will coincide. We simulated this
 mode and found a promising set of cuts to extract the $W\to jj$ signal. The
 basic cuts used for the $Zjj$ signal in Sec.~4 were adopted except that we
 required $p_T(Z) > 100\,\gev$, $p_T(jj) > 70\,\gev$ and $110 < Q = \MZjj -
 M_W - M_Z < 150\,\gev$. This removed some of the $\ta$ contribution for
 which the nominal $Q = 148\,\gev$. The mass distributions for $\sin\chi =
 1/3$ are shown in Fig.~\ref{fig:WZMdists} for $\int\CL dt = 20\,\ifb$. The
 LSTC signal more than doubles the number of standard model $W\to jj$ events
 in the $\Mjj$ distribution and it appears that the dijet signal should be
 observable with such a data set. Including the standard diboson events gives
 $S/\sqrt{B} = 4.0$ and $S/B = 0.08$ for $60 < \Mjj < 100\,\gev$. The $\MZjj$
 signal is problematic, but it may be possible to combine its significance
 with that for $\tro \to Z\tpi \to \ellp\ellm jj$. The $\dR$ and $\dX$
 distributions are in Fig.~\ref{fig:WZRXdists}. The narrow LSTC signal and
 the diboson contribution both peak very near $\dXm = 2\cos^{-1}(v_W) = 1.21$
 and they should be observable if the dijet excess is. The $\ellp\ellm jj$
 signal is only 60\% as large at $\sin\chi = 1/4$ as it is at $1/3$. It will
 be challenging to see it with $20\,\fb$ at $8\,\tev$.

\section*{Acknowledgments} 

We are grateful to K.~Black, T.~Bose, P.~Catastini, V.~Cavaliere, C.~Fantasia
and M.~Mangano for valuable conversations and advice. This work was
supported by Fermilab operated by Fermi Research Alliance, LLC,
U.S.~Department of Energy Contract~DE-AC02-07CH11359 (EE and AM) and in part
by the U.S.~Department of Energy under Grant~DE-FG02-91ER40676~(KL). KL's
research was also supported in part by Laboratoire d'Annecy-le-Vieux de
Physique Theorique (LAPTh) and the CERN Theory Group and he thanks LAPTh and
CERN for their hospitality.

\vfil\eject

\section*{Appendix: Nonanalytic Threshold Behavior of $d\sigma/d(\dR)$}

\subsection*{1. Kinematics}

We recall first the definition of the angles $\theta$, $\theta^*$, $\phi^*$
and the relevant coordinate systems. Choose the $z$-axis as the direction of
the incoming quark in the subprocess c.m.~frame (or the direction of the
harder initial-state parton in the $pp$ collision).  In the $\tro$ (or $\ta$)
rest frame, $\theta$ is the polar angle of the $\tpi$ velocity ${\bs v}$, the
angle it makes with the $z$-axis.  Define the $xz$-plane as the one
containing the unit vectors $\hat{\bs z}$ and $\hat{\bs v}$, so that
$\hat{\bs v} = \hat{\bs x}\sin\theta + \hat{\bs z}\cos\theta$, and $\hat{\bs
  y} = \hat{\bs z} \times \hat{\bs x}$. Define a starred coordinate system
{\em in the $\tpi$ rest frame} by making a rotation by angle~$\theta$ about
the $y$-axis of the $\tro$ frame. This rotation takes $\hat{\bs z}$ into
$\hat{\bs z}^* = \hat{\bs v}$ and $\hat{\bs x}$ into $\hat{\bs x}^* =
\hat{\bs x}\cos\theta - \hat{\bs z}\sin\theta$. In this frame, let $\hat {\bs
  p}_1^*$ be the unit vector in the direction of the jet (parton) making the
smaller angle with the direction of $\hat{\bs v}$.  This angle is $\theta^*$;
the azimuthal angle of ${\bs p}_1^* = -{\bs p}_2^*$ is $\phi^*$:
\be\label{eq:anglestwo}
\cos\theta = \hat{\bs z}\cdot \hat{\bs v}, \quad
\cos\theta^* = \hat{\bs p}_1^*\cdot \hat{\bs v}, \quad
\tan\phi^* = p_{1y^*}^*/p_{1x^*}^*.
\ee

The jets from $\tpi$ decay are labeled $j=1,2$ and they are assumed massless.
Let $\zeta_1 = +$ and $\zeta_2 = -$, and $\cth = \cos\theta$,
$\sth = \sin\theta$, etc. The boosted jets in the lab frame are
\bea\label{eq:jetmoms}
p_j^0 &=& \thalf M_{\tpi}\gamma(1+ \zeta_j v \cthst), \nn\\
{\bs p}_{j\parallel} &=& \thalf M_{\tpi}\gamma(v+ \zeta_j \cthst)(\hat{\bs
  x}\sth + \hat{\bs z}\cth), \nn\\ 
{\bs p}_{j\perp} &=& \thalf M_{\tpi} \zeta_j((\hat{\bs
  x}\cth - \hat{\bs z}\sth)\sthst\cphst + 
  \hat{\bs y}\sthst\sphst),
\eea
where $\gamma = (1-v^2)^{-\thalf}$.

We want to find the minimum of $\dR = \sqrt{(\deta)^2 + (\dphi)^2}$ as a
function of $\cth$, $\cthst$ and $\cphst$. From Eq.~(\ref{eq:jetmoms}),
\bea\label{eq:deta}
&&\deta = {\thalf}\ln\Biggl[
  \Bigl(\frac{1+v\cthst + (v+\cthst)\cth - \gamma^{-1} \sthst\cphst\sth}
             {1+v\cthst - (v+\cthst)\cth + \gamma^{-1}
               \sthst\cphst\sth}\Bigr)\nn\\
&&\qquad\qquad  \times\Bigl(\frac{1-v\cthst - (v-\cthst)\cth - \gamma^{-1}
  \sthst\cphst\sth} {1-v\cthst + (v-\cthst)\cth + \gamma^{-1}
               \sthst\cphst\sth}\Bigr) \Biggr]\,,
\eea
and
\bea\label{eq:cdphi}
 \cos(\dphi) &=& \frac{{\bs p}_{T1}\cdot {\bs p}_{T2}}{p_{T1}\,
  p_{T2}}\\
 &=& \frac{v^2 s^{2}_{\theta} - 
       \bigl(c^{2}_{\theta^{*}}\, s^{2}_{\theta} 
     + \gamma^{-2} s^{2}_{\theta^{*}}
        \bigl(c^{2}_{\theta}\, c^{2}_{\phi^{*}} +
        s^{2}_{\phi^{*}}\bigr)\bigr)
 - 2\gamma^{-1} s_{\theta^{*}} c_{\theta^{*}} \sth\, \cth\,
 c_{\phi^{*}}}
{\Bigl\{\bigl[v^2 s^2_{\theta} + \bigl(\cthst\sth +
  \gamma^{-1}\sthst\cphst\cth\bigr)^2 +  \bigl(\gamma^{-1}\sthst\sphst\bigr)^2
  \bigr]^2 - 4 v^2\sth^2 \bigl(\cthst\sth + \gamma^{-1}\sthst\cphst\cth
 \bigr)^2\Bigr\}^{1/2}}\,. \nn
\eea
%


\subsection*{2. Minimum of $\dR$}

It clearly is hopeless to deal with the analytic expression of $\Delta R$ as
a function of $\cth$, $\cthst$, $\cphst$. However, there is a simple way to
bypass it. The quantity
\be\label{eq:Delta}
\Delta \equiv \frac{M^2_{\tpi}}{2 p_{T1}\, p_{T2}} = \cosh(\deta) - \cos(\dphi)\,,
\ee
with $\deta \ge 0$ and $0 \le \dphi \le \pi$, is a monotonically increasing
function of $\Delta R$. This is seen by parametrizing
\be\label{eq:param}
\deta = \dR \cos\lambda\,, \quad \dphi = \dR\sin\lambda\,
\ee
with $\lambda \ge 0$ and $\lambda \le \pi/2$ if $\dR \le \pi$ or $\lambda
\le \sin^{-1}(\pi/\dR)$ if $\dR > \pi$. Then
\be\label{eq:diffparam}
\frac{\partial \Delta}{\partial(\dR)} = 
\cos\lambda \sinh(\deta) + \sin\lambda \sin(\dphi) \,.
\ee
This is non-negative. It vanishes only for (1) $\dR = 0$, which means $\Delta
=0$, and this cannot happen by its definition, Eq.~(\ref{eq:Delta}), and for
(2) $\deta = 0,\, \dphi = \pi$ meaning $\Delta R = \pi$; the latter is a
saddle point. This is the ``Col du Delta'', but it is one-sided, as shown in
Fig.~\ref{fig:CdD}.
\begin{figure}[!ht]
 \begin{center}
\includegraphics[width=3.15in, height=3.15in]{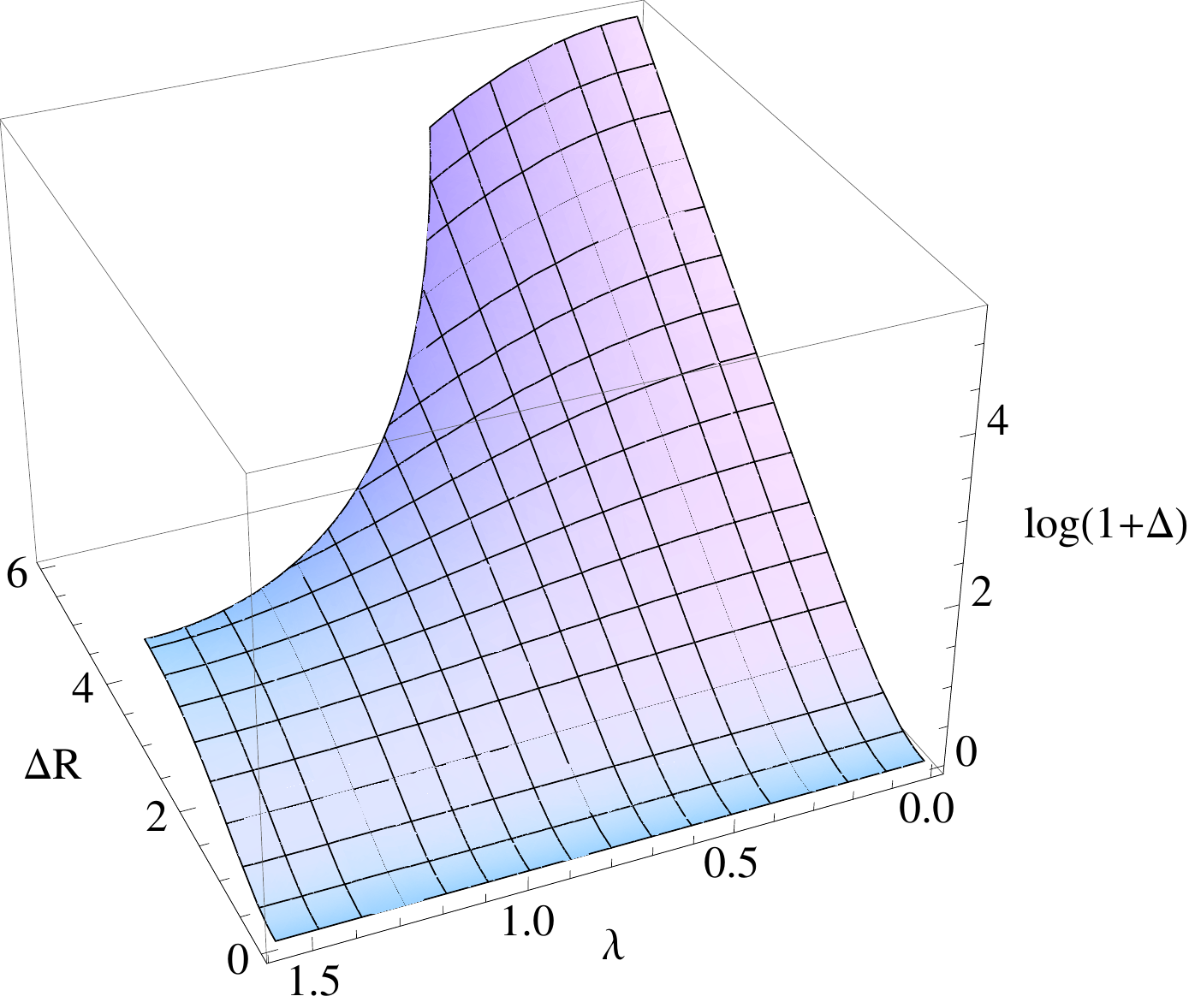}
\caption{The function $\ln(1 + \Delta)$ defined in
  Eqs.~(\ref{eq:Delta},\ref{eq:diffparam}). The Col du Delta at $\lambda =
  \pi/2$, $\dR = \pi$ is approached along the road $\lambda = \pi/2$. One
  cannot go over the pass and down the other side for the border is
  impassable. One must keep climbing along the ridge of increasing $\dR$ or
  return via the approach road.
  \label{fig:CdD}}
 \end{center}
 \end{figure}

 Minimizing $\Delta R$ thus amounts to minimizing $\Delta$, which in turn,
 amounts to maximizing $p_{T1}\,p_{T2}$. This is much simpler to examine than
 the original problem. We first maximize $p_{T1}\,p_{T2}$ at fixed $\cthst$,
 then maximize it with respect to $\cthst$. Since $p_{Tj} = \sqrt{p_{j0}^2 -
   p_{jz}^2}$ and $p_{j0}$ depends only on $\cthst$, $p_{T1}$ and $p_{T2}$
 are separately maximized at fixed $\cthst$ when $p_{1z} = p_{2z} = 0$. This
 requires $\cth = \sth\cphst = 0$. Then $p_{T1}\,p_{T2} = {(\thalf} \gamma
 M_{\tpi})^2 (1 - v^2\cos^2\theta^*)$ is maximized at $\cthst = 0$. In
 conclusion, $\dR$ is minimized if and only if
\be\label{eq:dRmin}
\cth = \cthst = \cphst = 0\,.
\ee
This corresponds to two distinct, isolated points in the angular phase space
($\phi^{*} = \pi/2, \, 3\pi/2$). The degeneracy of the minimum is only
discrete. At $\dR$'s minimum, $\deta = 0$ and $\dphi = \cos^{-1}(2v^2 - 1) =
2\cos^{-1}(v) \equiv \dXm$, so that
\be\label{eq:dRm}
\dRm = \dXm = 2\cos^{-1}(v)\,.
\ee

\subsection*{3. Local behavior around $\cos\theta= \cos\theta^* = \cos\phi^*  = 0$}

We now investigate the behavior of $\dR$ as a function of 
$\cth, \cphst$ and $\cthst$ around its minimum at $\cth = \cthst = \cphst =
0$ by means of a Taylor expansion of at most second order in any of
these variables. From, Eqs.~(\ref{eq:deta},\ref{eq:cdphi}), we obtain
\bea\label{eq:Taylor}
(\deta)^2 &=& 4 \gamma^{-2} \cphst^2 + \CO(c^3)\,,\\
\cos(\dphi) &=& \cos\dXm - (1 - \cos\dXm)\, v^2(\cth^{2} +
\cthst^2) + (1 + \cos\dXm)\, \gamma^{-2}\cphst^2 +
\CO(c^3)\,.\nn
\eea
Interpreting the latter equation as:
\be\label{eq:Taylortwo}
\cos(\dphi) = \cos\dXm - \sin\dXm\,(\dphi - \dXm) + \CO((\dphi - \dXm)^2)
\ee
we identify
\be\label{eq:dphidXm}
\dphi = \dXm + \left[v^2\tan(\dXm/2)\,(\cth^{2} + \cthst^2) - 
 \gamma^{-2}\cot(\dXm/2) c_{\phi^{*}}^{2} + \CO(c^3)\right]\,. 
\ee
Then
\be\label{eq:dRbexp}
\dR \equiv \sqrt{(\deta)^2 + (\dphi)^2} = \dXm + {\thalf}
\left(\bth\cth^2 + \bthst\cthst^2 + \bphst\cphst^2\right) +\CO(c^3)\,,
\ee
where
\bea\label{eq:bterms}
\bth &=& \bthst = 2 v^2\tan(\dXm/2) = 2v\gamma^{-1}\,,\nn\\
\bphst &=& 2\gamma^{-2}\bigl(2/\dXm - v\gamma\bigr)\,.
\eea

The shape of the surface $\Delta R = f(\cth,\cthst,\cphst)$ in the
neighborhood of the minimum $\Delta R =\Delta \chi_{min}$
is a convex paraboloid with ellipsoidal section whose eigen-directions are
parallel to the axes of the coordinates $\cth$, $\cthst$ and $\cphst$.  The
curvature is $> 0$ along each of these axes for all $0 < v < 1$; i.e. there
is no flat direction, as expected from the fact the minimum is at isolated
point(s).

\subsection*{4. Calculation of the singular part of  $d\sigma/d(\Delta R)$}

The differential cross section for $\bar q q \to \tro,\ta \to W/Z\tpi$,
followed by $\tpi \to \bar q q$ is~\footnote{Since there are two points in
  the $(\cth,\cthst,\cphst)$ phase space where $\dR$ has a minimum, $\theta
  = \theta^{*}= \pi/2$ and $\phi^{*} = \pi/2, \, 3\pi/2$, it is more
  convenient to use the variable $\cphst$ instead of $\phi^*$. This
  introduces (a) the Jacobian $(1-\cphst^2)^{-1/2}$ which is one at $\cphst =
  0$; and (b) a factor of two to account for the contributions of the two
  minima in the calculation of the normalization coefficient.}
\be\label{eq:dsig} 
d\sigma =\left[\frac{d\sigma(\bar qq \to W/Z\tpi)}{d\cth}\right] \, B(\pi_T
\to \bar qq)) 
\underbrace{
\left[\frac{1}{\Gamma(\pi_T \to \bar qq)} 
\frac{d\Gamma(\pi_T \to \bar q q)}{d\cthst\,d\cphst} \right]
}_{\left(2\pi\sqrt{1-\cphst^2}\right)^{-1}}
\,{d\cth\, d\cthst\,d\cphst} \,.
\ee
To compute the distribution in a compound variable $\zeta$, such as $\Delta
\chi$ or $\Delta R$, we use a Fadeev-Popov-like trick
\be\label{eq:FP}
1 = \int d\zeta \,\delta\left(\zeta - f\left(\cth,\cthst,\cphst\right)\right)\,.
\ee
where $f(\cth,\cthst,\cphst)$ gives the expression of $\zeta$ in terms of
the phase space variables.  The $\zeta$-distribution is then
\bea\label{eq:zetadist}
\frac{d\sigma}{d\zeta} &=&
\int d\sigma(\mbox{from Eq.~(\ref{eq:dsig})}) \,
\delta\left(\zeta - f\left(\cth,\cthst\cphst\right)\right)\,.
\eea
Let $\zeta = \Delta R$ be slightly above and close to $\dXm$, and define $\omega
= \Delta R - \dXm$ to shorten expressions. Solving Eq.~(\ref{eq:dRbexp}) with
respect to $\cthst$ gives
%
\be\label{eq:cthstsolns}
\cthst = \pm \hat{c}_{\theta^*} = \pm\sqrt{\biggl(\frac{2}{\bthst}\biggr) 
\biggl(\omega - {\thalf}\bigl(\bth\cth^2 + \bphst\cphst^2\bigr)
+\CO(c^3)\biggr)}\,.
\ee
Notice that Eq.~(\ref{eq:cthstsolns}) has to be supplemented by the
restriction
\be\label{eq:restrict}
\omega - {\thalf}(\bth\cth^2 + \bphst\cphst^2 +\CO(c^3)) \ge 0\,.
\ee
%
Substituting
\be\label{eq:subs}
\delta(\Delta R - f(\cth,\cthst,\cphst))
= (\bthst\hat c_{\theta^*})^{-1} \, \left[ 
 \delta \left( c_{\theta^{*}} - \hat{c}_{\theta^{*}}\right)
 +
 \delta \left( c_{\theta^{*}} + \hat{c}_{\theta^{*}}\right) \right] 
\Theta 
 \left[
   \omega 
   - \frac{1}{2}
   \left( 
   b_{\theta} \cth^{2} +  
   b_{\phi^{*}} c_{\phi^{*}}^{2} + o(c_{j}^{3})
  \right)
 \right]
\ee
in Eq.~(\ref{eq:FP}) and integrating over $\cthst$ leads to the following
threshold behavior for the cross section:
\bea\label{eq:dsigdRtwo}
\left(\frac{d\sigma}{d(\dR)}\right)_{\rm threshold} &\simeq&
\left[\frac{d\sigma(\bar qq \to W/Z\tpi)}{d\cth}\right]_{\cth=\cthst\cphst=0}
 \,B(\tpi \to \bar qq)\\
&&\times \frac{\sqrt{2}}{2\pi} \,
\left(\frac{1}{\bthst}\right)^{1/2}
\int d\cth d\cphst
\frac{\Theta 
 \left[
   \omega  - \frac{1}{2}\left(\bth\cth^2 + \bphst\cphst^2 + \CO(c^3)\right)
 \right]}
{\left[
   \omega  - \frac{1}{2}\left(\bth\cth^2 + \bphst\cphst^2 + \CO(c^3)\right)
 \right]^{1/2}}\,.  \nn
\eea
It is convenient to trade $\cth, \cphst$ for new variables $\rho,\kappa$:
\be\label{eq:tradevars}
\rho\cos\kappa = \sqrt{\bth/2}\,\cth\,,\quad
\rho\sin\kappa = \sqrt{\bphst/2}\,\cphst\,,\qquad
(0 \le \rho \le \sqrt{\omega}\,, \quad 0 \le \kappa < 2\pi)\,.
\ee
The integral in Eq.~(\ref{eq:dsigdRtwo}) then yields our final result, the
square-root behavior of $d\sigma/d(\dR)$ at threshold:
\be\label{eq:dsigdRfinal}
\left(\frac{d\sigma}{d(\dR)}\right)_{\rm threshold} \simeq
2^{3/2}\sqrt{\frac{\dR - \dXm}{\bth\,\bthst\,\bphst}}
\left[
\frac{d\sigma(\bar qq \to W/Z\tpi)}{d\cth}\right]_0 \,B(\tpi \to \bar qq)\,.
\ee
\vfil\eject

\bibliography{TC_at_LHC_8TeV}

\providecommand{\href}[2]{#2}\begingroup\raggedright\begin{thebibliography}{10}

\bibitem{Aaltonen:2011mk}
{\bf CDF} Collaboration, T.~Aaltonen {\em et.~al.}, ``{Invariant Mass
  Distribution of Jet Pairs Produced in Association with a W boson in ppbar
  Collisions at sqrt(s) = 1.96 TeV},'' {\em Phys. Rev. Lett.} {\bf 106} (2011)
  171801, \href{http://xxx.lanl.gov/abs/1104.0699}{ 1104.0699}.

\bibitem{CDFnew}
{\bf CDF} Collaboration
  {http://www-cdf.fnal.gov/physics/ewk/2011/wjj/7$\textunderscore$3.html}.

\bibitem{Abazov:2011af}
{\bf D0} Collaboration, V.~M. Abazov, ``{Bounds on an anomalous dijet resonance
  in W+jets production in ppbar collisions at sqrt{s} =1.96 TeV},'' {\em Phys.
  Rev. Lett.} {\bf 107} (2011) 011804,
  \href{http://xxx.lanl.gov/abs/1106.1921}{ 1106.1921}.

\bibitem{AnnoviLP11}
A.~Annovi, ``{Physics Beyond the Standard Model, talk at Lepton-Photon 2011,
  Mumbai, India}.'' 2011.

\bibitem{Eichten:2011sh}
E.~J. Eichten, K.~Lane, and A.~Martin, ``{Technicolor at the Tevatron},'' {\em
  Phys. Rev. Lett.} {\bf 106} (2011) 251803,
  \href{http://xxx.lanl.gov/abs/1104.0976}{ 1104.0976}.

\bibitem{Eichten:2011xd}
E.~Eichten, K.~Lane, and A.~Martin, ``{Testing CDF's Dijet Excess and
  Technicolor at the LHC},'' \href{http://xxx.lanl.gov/abs/1107.4075}{
  1107.4075}.

\bibitem{Eichten:2012br}
E.~Eichten, K.~Lane, A.~Martin, and E.~Pilon, ``{Testing the Technicolor
  Interpretation of CDF's Dijet Excess at the LHC},''
  \href{http://xxx.lanl.gov/abs/1201.4396}{ 1201.4396}.

\bibitem{Holdom:1981rm}
B.~Holdom, ``Raising the sideways scale,'' {\em Phys. Rev.} {\bf D24} (1981)
  1441.

\bibitem{Appelquist:1986an}
T.~W. Appelquist, D.~Karabali, and L.~C.~R. Wijewardhana, ``Chiral hierarchies
  and the flavor changing neutral current problem in technicolor,'' {\em Phys.
  Rev. Lett.} {\bf 57} (1986) 957.

\bibitem{Yamawaki:1986zg}
K.~Yamawaki, M.~Bando, and K.-i. Matumoto, ``Scale invariant technicolor model
  and a technidilaton,'' {\em Phys. Rev. Lett.} {\bf 56} (1986) 1335.

\bibitem{Akiba:1986rr}
T.~Akiba and T.~Yanagida, ``Hierarchic chiral condensate,'' {\em Phys. Lett.}
  {\bf B169} (1986) 432.

\bibitem{Eichten:1979ah}
E.~Eichten and K.~D. Lane, ``{Dynamical Breaking of Weak Interaction
  Symmetries},'' {\em Phys. Lett.} {\bf B90} (1980) 125--130.

\bibitem{Lane:1989ej}
K.~D. Lane and E.~Eichten, ``{Two Scale Technicolor},'' {\em Phys. Lett.} {\bf
  B222} (1989) 274.

\bibitem{Abazov:2006iq}
{\bf D0 Collaboration} Collaboration, V.~Abazov {\em et.~al.}, ``{Search for
  techniparticles in e+jets events at D0},'' {\em Phys.Rev.Lett.} {\bf 98}
  (2007) 221801, \href{http://xxx.lanl.gov/abs/hep-ex/0612013}{
  hep-ex/0612013}.

\bibitem{Aaltonen:2009jb}
{\bf CDF Collaboration} Collaboration, T.~Aaltonen {\em et.~al.}, ``{Search for
  Technicolor Particles Produced in Association with a W Boson at CDF},'' {\em
  Phys.Rev.Lett.} {\bf 104} (2010) 111802,
  \href{http://xxx.lanl.gov/abs/0912.2059}{ 0912.2059}.

\bibitem{Lane:2009ct}
K.~Lane and A.~Martin, ``{An Effective Lagrangian for Low-Scale Technicolor},''
  {\em Phys. Rev.} {\bf D80} (2009) 115001,
  \href{http://xxx.lanl.gov/abs/0907.3737}{ 0907.3737}.

\bibitem{Lane:2002sm}
K.~Lane and S.~Mrenna, ``{The collider phenomenology of technihadrons in the
  technicolor Straw Man Model},'' {\em Phys. Rev.} {\bf D67} (2003) 115011,
  \href{http://xxx.lanl.gov/abs/hep-ph/0210299}{ hep-ph/0210299}.

\bibitem{Eichten:2007sx}
E.~Eichten and K.~Lane, ``{Low-scale technicolor at the Tevatron and LHC},''
  {\em Phys.Lett.} {\bf B669} (2008) 235--238,
  \href{http://xxx.lanl.gov/abs/0706.2339}{ 0706.2339}.

\bibitem{Brooijmans:2008se}
G.~Brooijmans {\em et.~al.}, ``{New Physics at the LHC: A Les Houches Report.
  Physics at Tev Colliders 2007 -- New Physics Working Group},''
  \href{http://xxx.lanl.gov/abs/0802.3715}{ 0802.3715}.

\bibitem{Brooijmans:2010tn}
G.~Brooijmans {\em et.~al.}, ``{New Physics at the LHC. A Les Houches Report:
  Physics at TeV Colliders 2009 - New Physics Working Group},''
  \href{http://xxx.lanl.gov/abs/1005.1229}{ 1005.1229}.

\bibitem{Sjostrand:2006za}
T.~Sjostrand, S.~Mrenna, and P.~Skands, ``PYTHIA 6.4 physics and manual,'' {\em
  JHEP} {\bf 05} (2006) 026, \href{http://xxx.lanl.gov/abs/hep-ph/0603175}{
  hep-ph/0603175}.

\bibitem{Kilic:2011sr}
C.~Kilic and S.~Thomas, ``{Signatures of Resonant Super-Partner Production with
  Charged-Current Decays},'' {\em Phys. Rev.} {\bf D84} (2011) 055012,
  \href{http://xxx.lanl.gov/abs/1104.1002}{ 1104.1002}.

\bibitem{Cao:2011yt}
Q.-H. Cao {\em et.~al.}, ``{W plus two jets from a quasi-inert Higgs
  doublet},'' {\em JHEP} {\bf 08} (2011) 002,
  \href{http://xxx.lanl.gov/abs/1104.4776}{ 1104.4776}.

\bibitem{Chen:2011wp}
C.-H. Chen, C.-W. Chiang, T.~Nomura, and Y.~Fusheng, ``{A light charged Higgs
  boson in two-Higgs doublet model for CDF $Wjj$ anomaly},''
  \href{http://xxx.lanl.gov/abs/1105.2870}{ 1105.2870}.

\bibitem{Fan:2011vw}
J.~Fan, D.~Krohn, P.~Langacker, and I.~Yavin, ``{A Higgsophilic s-channel Z'
  and the CDF W+2J Anomaly},'' {\em Phys. Rev.} {\bf D84} (2011) 105012,
  \href{http://xxx.lanl.gov/abs/1106.1682}{ 1106.1682}.

\bibitem{Ghosh:2011np}
D.~K. Ghosh, M.~Maity, and S.~Roy, ``{R parity violating supersymmetric
  explanation for the CDF Wjj excess},'' {\em Phys. Rev.} {\bf D84} (2011)
  035022, \href{http://xxx.lanl.gov/abs/1107.0649}{ 1107.0649}.

\bibitem{Gunion:2011bx}
J.~F. Gunion, ``{A two-Higgs-doublet interpretation of a small Tevatron $Wjj$
  excess},'' \href{http://xxx.lanl.gov/abs/1106.3308}{ 1106.3308}.

\bibitem{Buckley:2011vc}
M.~R. Buckley, D.~Hooper, J.~Kopp, and E.~Neil, ``{Light Z' Bosons at the
  Tevatron},'' {\em Phys. Rev.} {\bf D83} (2011) 115013,
  \href{http://xxx.lanl.gov/abs/1103.6035}{ 1103.6035}.

\bibitem{Hewett:2011fk}
J.~Hewett and T.~Rizzo, ``Dissecting the Wjj Anomaly: Diagnostic Tests of a
  Leptophobic Z','' \href{http://xxx.lanl.gov/abs/1106.0294v1}{ 1106.0294v1}.

\bibitem{Harnik:2011mv}
R.~Harnik, G.~D. Kribs, and A.~Martin, ``{Quirks at the Tevatron and Beyond},''
  {\em Phys. Rev.} {\bf D84} (2011) 035029,
  \href{http://xxx.lanl.gov/abs/1106.2569}{ 1106.2569}.

\bibitem{Dobrescu:2011fk}
B.~A. Dobrescu and G.~Z. Krnjaic, ``Weak-triplet, color-octet scalars and the
  CDF dijet excess,'' \href{http://xxx.lanl.gov/abs/1104.2893v1}{ 1104.2893v1}.

\bibitem{Nelson:2011us}
A.~E. Nelson, T.~Okui, and T.~S. Roy, ``{A unified, flavor symmetric
  explanation for the t-tbar asymmetry and Wjj excess at CDF},'' {\em Phys.
  Rev.} {\bf D84} (2011) 094007, \href{http://xxx.lanl.gov/abs/1104.2030}{
  1104.2030}.

\bibitem{Yu:2011cw}
F.~Yu, ``{A Z' Model for the CDF Dijet Anomaly},'' {\em Phys. Rev.} {\bf D83}
  (2011) 094028, \href{http://xxx.lanl.gov/abs/1104.0243}{ 1104.0243}.

\bibitem{Cheung:2011zt}
K.~Cheung and J.~Song, ``{Baryonic Z' Explanation for the CDF Wjj Excess},''
  {\em Phys. Rev. Lett.} {\bf 106} (2011) 211803,
  \href{http://xxx.lanl.gov/abs/1104.1375}{ 1104.1375}.

\bibitem{Eichten:1997yq}
E.~Eichten, K.~D. Lane, and J.~Womersley, ``{Finding low scale technicolor at
  hadron colliders},'' {\em Phys.Lett.} {\bf B405} (1997) 305--311,
  \href{http://xxx.lanl.gov/abs/hep-ph/9704455}{ hep-ph/9704455}.

\bibitem{ATLASWjj}
{\bf ATLAS} Collaboration, ``Invariant mass distribution of jet pairs produced
  in association with a leptonically decaying $W$ boson using $1.02\,{\rm
  fb}^{-1}$ of ATLAS data.'' ATLAS-CONF-2011-097.

\bibitem{Mangano:2002ea}
M.~L. Mangano, M.~Moretti, F.~Piccinini, R.~Pittau, and A.~D. Polosa, ``ALPGEN,
  a generator for hard multiparton processes in hadronic collisions,'' {\em
  JHEP} {\bf 07} (2003) 001, \href{http://xxx.lanl.gov/abs/hep-ph/0206293}{
  hep-ph/0206293}.

\bibitem{MLM}
M.~Mangano http://mlm.web.cern.ch/mlm/talks/lund-alpgen.pdf, 2004.

\bibitem{Campbell:2011bn}
J.~M. Campbell, R.~K. Ellis, and C.~Williams, ``{Vector boson pair production
  at the LHC},'' {\em JHEP} {\bf 07} (2011) 018,
  \href{http://xxx.lanl.gov/abs/1105.0020}{ 1105.0020}.

\bibitem{Cacciari:2005hq}
M.~Cacciari and G.~P. Salam, ``{Dispelling the $N^{3}$ myth for the $k_t$
  jet-finder},'' {\em Phys. Lett.} {\bf B641} (2006) 57--61,
  \href{http://xxx.lanl.gov/abs/hep-ph/0512210}{ hep-ph/0512210}.

\bibitem{Campbell:2011gp}
J.~M. Campbell, A.~Martin, and C.~Williams, ``{NLO predictions for a lepton,
  missing transverse momentum and dijets at the Tevatron},'' {\em Phys.Rev.}
  {\bf D84} (2011) 036005, \href{http://xxx.lanl.gov/abs/1105.4594}{
  1105.4594}.

\bibitem{Frederix:2011ig}
R.~Frederix, S.~Frixione, V.~Hirschi, F.~Maltoni, R.~Pittau, {\em et.~al.},
  ``{aMC@NLO predictions for Wjj production at the Tevatron},'' {\em JHEP} {\bf
  1202} (2012) 048, \href{http://xxx.lanl.gov/abs/1110.5502}{ 1110.5502}.

\bibitem{CMSWjj}
{\bf CMS} Collaboration, ``Study of the dijet invariant mass distribution in $W
  \to \ell\nu$ plus jets events produced in pp collisions at $\sqrt{s} =
  7\,{\rm TeV}$.'' CMS PAS EWK-11-017
  [https://cdsweb.cern.ch/record/1431015/files/EWK-11-017-pas.pdf].

\bibitem{CMSWZ}
{\bf CMS} Collaboration, ``Search for exotic particles decaying to the WZ final
  state with the CMS Experiment.''
  {https://twiki.cern.ch/twiki/bin/view/CMSPublic/PhysicsResultsEXO11041Winter%
2012}.

\bibitem{Eichten:1984eu}
E.~Eichten, I.~Hinchliffe, K.~D. Lane, and C.~Quigg, ``{Super Collider
  Physics},'' {\em Rev.Mod.Phys.} {\bf 56} (1984) 579--707.

\bibitem{Abazov:2009eu}
{\bf D0} Collaboration, V.~M. Abazov {\em et.~al.}, ``{Search for a resonance
  decaying into WZ boson pairs in $p\bar{p}$ collisions},'' {\em Phys. Rev.
  Lett.} {\bf 104} (2010) 061801, \href{http://xxx.lanl.gov/abs/0912.0715}{
  0912.0715}.

\end{thebibliography}\endgroup
\bibliographystyle{utcaps}
\end{document}